\documentclass[a4paper]{article}  
\usepackage{fullpage}
\usepackage{graphicx,latexsym,amsfonts,amsmath,amssymb,verbatim,amsmath,subfigure,fancyhdr,multicol}
\usepackage{epsfig}
\usepackage{color}
\usepackage{hyperref}

\usepackage[english]{babel}
\usepackage[utf8]{inputenc}
\usepackage[T1]{fontenc}
\usepackage{times}

\usepackage{fancyhdr}

\usepackage{listings}

\usepackage{multicol}

\lstdefinestyle{code}
{
frame=single, 
columns=fullflexible,
rulesepcolor=\color{black}, 
keywordstyle=\color{vert},
basicstyle=\scriptsize, 
}

\definecolor{orange}{rgb}{0.8,0.4,0.3}
\definecolor{vert}{rgb}{0,0.3,0}

\lstdefinestyle{c}
{style=code, language=c, tabsize=2,
morecomment=[s][\color{orange}]{/*}{*/},
morecomment=[s][\color{orange}]{/**}{*/},
moredelim=[is][\color{red}]{/*!\ }{*/},
morecomment=[l][\color{orange}]{//},
moredelim=**[is][\only<1>{\pause}]{//pause>}{//<pause},
}

\lstdefinestyle{java}
{style=code, language=java, tabsize=2,
morecomment=[s][\color{orange}]{/*}{*/},
morecomment=[s][\color{orange}]{/**}{*/},
moredelim=[is][\color{red}]{/*!\ }{*/},
morecomment=[l][\color{orange}]{//},
moredelim=**[is][\only<1>{\pause}]{//pause>}{//<pause}
}

\lstdefinestyle{python}
{
style=code, language=python,
morecomment=[l][\color{orange}]{\#}
}

\lstdefinestyle{ruby}
{
style=code, language=ruby,
morecomment=[l][\color{orange}]{\#},
}

\lstdefinestyle{xml}
{
style=code, language=XML,
moredelim=[s][\color{vert}]{<}{>},
}

\lstdefinestyle{bash}
{basicstyle=\small\ttfamily, frame=lines, rulesepcolor=\color{black},
language=bash,
columns=fullflexible,
}

\lstdefinestyle{italique}
{basicstyle=\itshape\ttfamily,identifierstyle=\itshape}

\newcommand{\bash}{\lstinline[language=bash,style=italique]}

\newcommand{\jv}{\lstinline[language=java,style=italique]}

\begin{document}


%
\thispagestyle{empty}\vspace*{-1.5cm}\centerline{
\epsfig{file=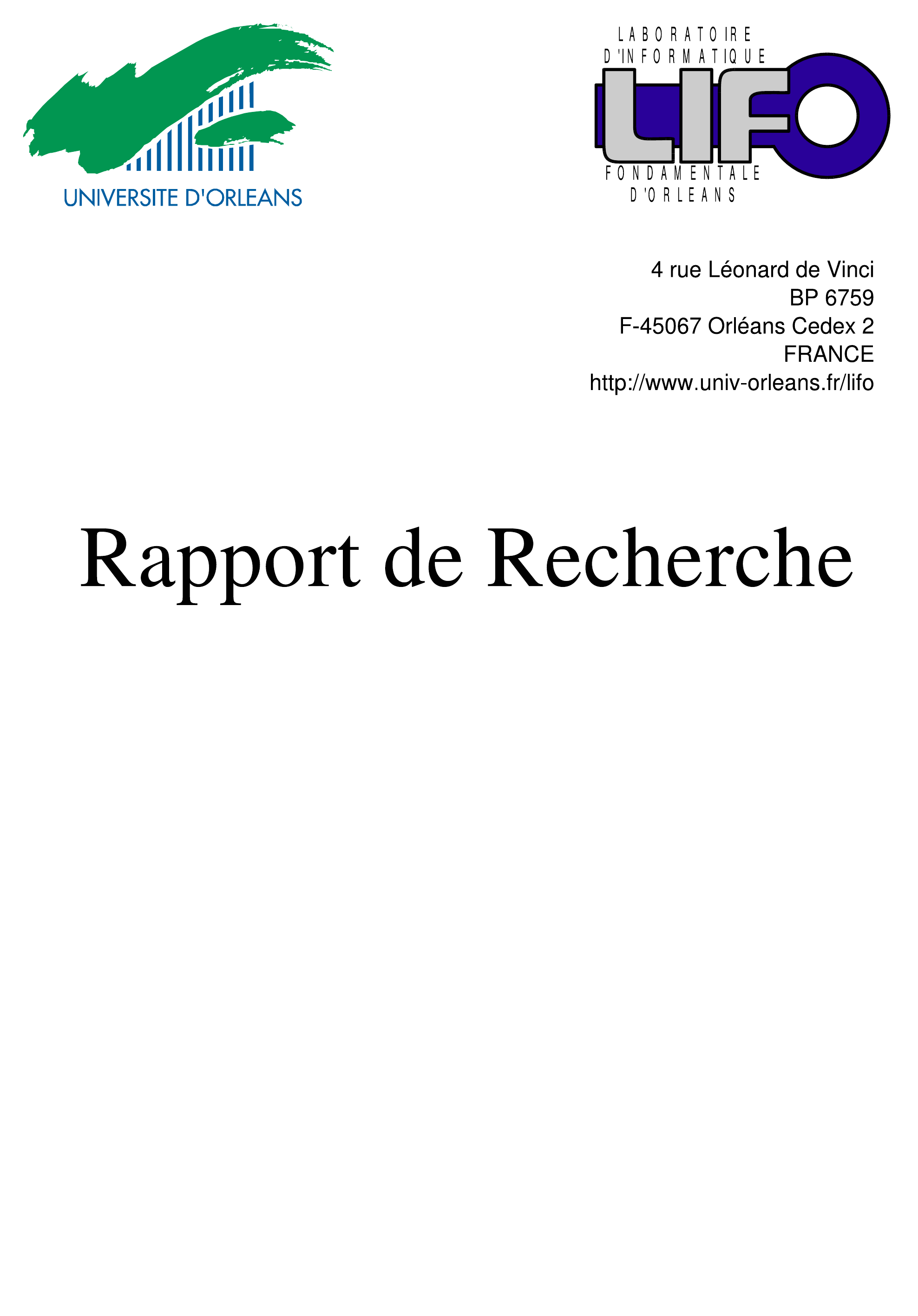,height=27cm}}
\vspace*{-13.5cm}\hspace*{4cm}\begin{minipage}[b]{10.7cm}{\Huge
\begin{flushright}
%
%
{\bf Implementation of exponential and parametrized algorithms in the AGAPE project}
\end{flushright}}\vspace*{0.5cm}
%
%
{\Large\begin{flushright}
P. Berthom\'e, J.-F.~Lalande, V.~Levorato\\[1mm]
LIFO, Universit\'e d'Orl\'eans - ENSI de Bourges
\end{flushright}}
\vspace{1mm}
%
%
{\Large\begin{flushright}
Report n$^o$ {\bf RR-2012-01}\\
Version 1
\end{flushright}}\end{minipage}\newpage
%
%

\pagestyle{fancy}
\newpage

\tableofcontents
\newpage

%

\addcontentsline{toc}{section}{Presentation}  
\section*{Presentation}

This technical report describes the implementation of exact and parametrized exponential algorithms, developed during the French \href{http://www-sop.inria.fr/mascotte/Contrats/AGAPE/}{ANR Agape} during 2010-2012. The developed algorithms are distributed under the CeCILL license and have been written in Java using the \href{http://jung.sourceforge.net/}{Jung} graph library.

\section{Installation notes}

The source code is available at \url{http://traclifo.univ-orleans.fr/Agape/}. Several dependencies are necessary in order to use the Agape library. They are included in the \bash{lib/} directory of the SVN trunk:

\begin{itemize}
 \item \href{http://jung.sourceforge.net/}{Jung library 2.0.1}, BSD license (\bash{jung*.jar}).
 \item \href{http://larvalabs.com/collections/}{Apache commons for collections}, Apache License 2.0 (\bash{collections-generic-4.01.jar})\footnote{This is a new (very old i.e. 2006) version of the popular Jakarta Commons-Collections project. It features support for Java 1.5 Generics. Generics introduce a whole new level of usability and type-safety to the Commons-Collections classes.}.
 \item \href{http://code.google.com/p/guava-libraries/}{Guava i.e. Google's collections}, Apache License 2.0, (\bash{guava-r09.jar}).
 \item \href{http://www-sop.inria.fr/mascotte/mascopt/}{Mascopt optimization library}, LGPL, (\bash{mascoptLib.jar}).
\end{itemize}

The project can be downloaded as an Eclipse project from the SVN trunk repository (see instructions on the website). It is particularly useful
for studying the source code of the provided algorithms. To use this library as a dependency in another project, only the agape.jar file and the dependencies (dependencies.zip) are needed.

\section{Packages content}  

In this section, we briefly describe the implemented algorithms. Section~\ref{sec:tutorials} describes how to use these algorithms through simple tutorials.

\subsection{Package agape.algos}  

This package contains the different implemented exponential or parametrized algorithms. The abstract class \jv{Algorithms} groups the different \emph{factories} for graph/vertices/arcs creation. The graphs are based on generic classes of the Jung library. Indeed, some algorithms require to instantiate vertices or arcs. The instanciation of such objects depends of the nature of the generic token, e.g. a vertex can be an \jv{Integer}. Thus, as the type of vertices are not known in advance, the algorithms that require to create graphs should obtain a factory for the provided graph.

\begin{figure}[bH]
\begin{center}
\includegraphics[width=3.5in]{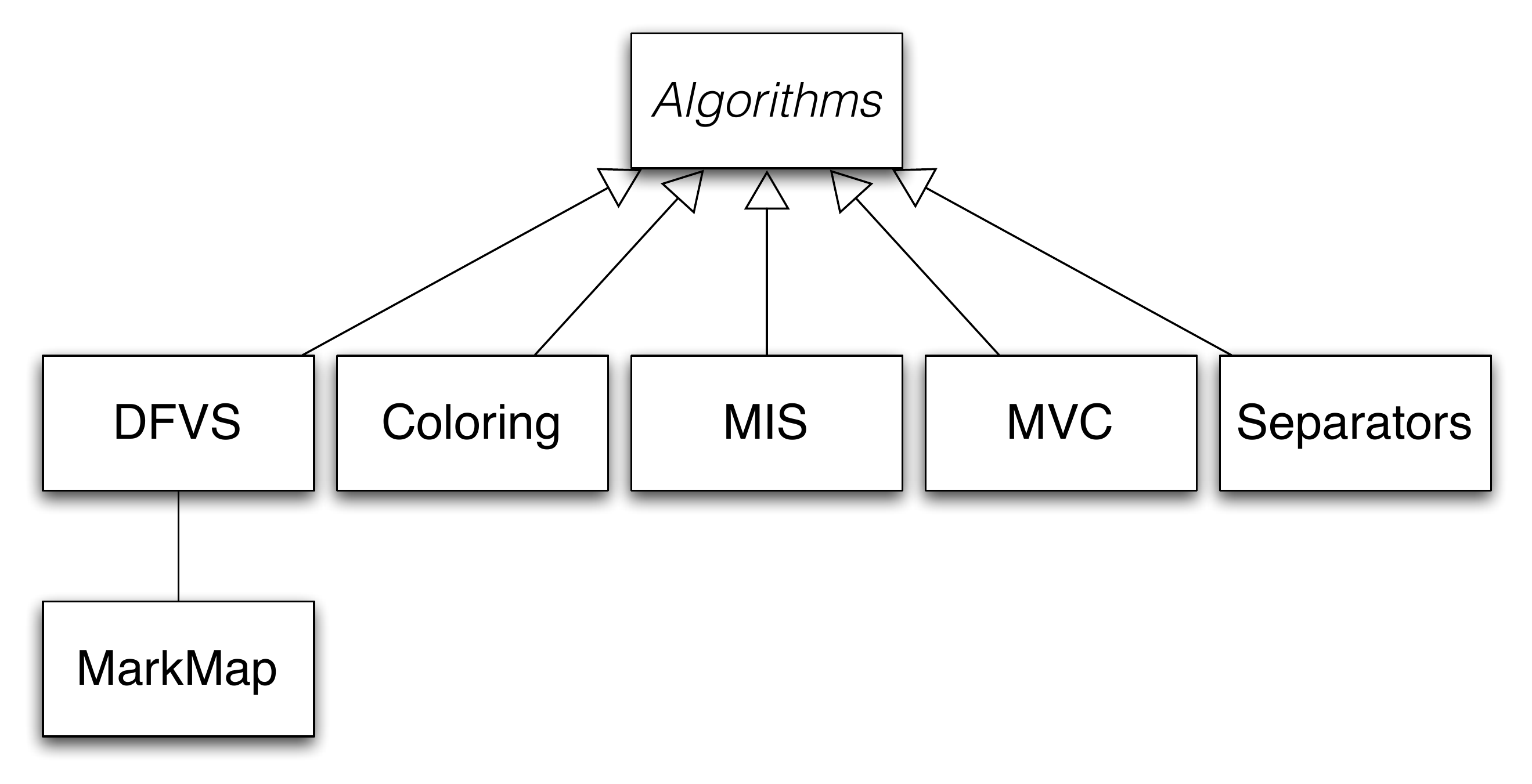}
\caption{The \emph{agape.algos} package hierarchy}
\label{fig:algos}
\end{center}
\end{figure}

The different classes and their hierarchy are represented in Figure~\ref{fig:algos}. Table~\ref{tab:algos} sums-up the different implemented algorithms, their complexity in time and the references to the papers that provided the algorithms. The last column recalls the acronym used for the Agape Command Line tool (c.f. Section~\ref{sec:CL}). 
In the following subsections, each implemented algorithm is briefly described.

\begin{table}
\begin{center}
\begin{tabular}{|lllll|}
\hline
Acronym & Algorithm & Time complexity & References & Agape CL\\
\hline
\hline
CN   & Chromatic Number & & & \\
     & ~~~\emph{Bodlaender, Kratsch} & $O(5.283^n)$ & \cite{Bodlaender2006} & CN\\
\hline
DFVS & Directed Feedback Vertex Set & & & \\
     & ~~~\emph{Razgon, Thomass\'e} & $O(1.9977^n)$ & \cite{Razgon2007, Thomasse2010} & DFVS\\
\hline
MIS & Maximum Independent Set & & & \\
     & ~~~\emph{Brute-Force}                   &  $O(n^2.2^n)$ & & MISBF \\
     & ~~~\emph{Moon, Moser}&	 $O(1.4423^n)$ & \cite{Moon1965} & MISMM \\
     & ~~~\emph{Fomin, Grandoni, Kratsch} & $O(1.2201^n)$ & \cite{Fomin2009} & MISFGK\\
\hline
MVC & Minimum Vertex Cover & & & \\
     & ~~~\emph{Brute-Force}                   & $O( {n \choose k})$ & & MVCBF\\
     & ~~~\emph{Niedermeier}&  $O(1.47^k)$ & \cite{Niedermeier2006a} & MVCDBS\\
     & ~~~\emph{Niedermeier}&  $O(1.33^k)$ & \cite{Niedermeier2006a} & MVCN\\
     & ~~~\emph{Buss, Goldsmith}&  -- & \cite{Niedermeier2006a,Buss1993} & MVCBG\\

\hline
SEP & Minimal ab-Separators & & &\\
     & ~~~\emph{Berry, Bordat, Cogis} & $O(n^3)$ & \cite{Berry1999} & SEP\\
\hline
\end{tabular}
\end{center}
\caption{Algorithms summary}
\label{tab:algos}
\end{table}

\subsubsection{Class Coloring}

The goal of this class is to compute the minimum number of colors to properly color an undirected graph.

\begin{description}
\item[int chromaticNumber(Graph<V,E> G):] Computes the chromatic number of a graph G. This method is based on the
\emph{An exact algorithm for graph coloring with polynomial memory} by Bodlaender and Kratsch~\cite{Bodlaender2006}.
The algorithm solves the problem in PSPACE and in time $O(5.283^n)$.
\item[Set<Set<V> > graphColoring(Graph<V,E> G):] Computes a solution having the minimum chromatic number using the same algorithm as before. It returns the partition of the vertices into color classes.
\item[Set<Set<V> > greedyGraphColoring(Graph<V,E> G):] a greedy algorithm that computes iteratively the maximum independet set of G and removes it.
\end{description}

\subsubsection{Class MinDFVS}

This class is dedicated to the minimum directed feedback vertex set. This problem consists in finding the minimum number of vertices to delete in order to get a directed acyclic graph from a directed graph.

\begin{description}
\item[Set<V> minimumDirectedAcyclicSubset(Graph<V,E> G):] Computes the Minimum Directed Acyclic Subset as the complement of the Maximum Directed Feedback Vertex Set (see below).
\item[Set<V> maximumDirectedAcyclicSubset(Graph<V,E> G):]
The implemented method solves the problem in time $O(1.9977^n)$ using the Razgon
     work \emph{Computing Minimum Directed Feedback Vertex Set in $O*(1.9977^n)$}~\cite{Razgon2007}). It uses kernelization techniques presented in \emph{A $4k^2$ kernel for  feedback vertex set} by Thomassé~\cite{Thomasse2010}.
\item[Set<V> greedyMinFVS(Graph<V,E> G):] Computes an approximation of the FVS problem using the big degree heuristic, written by Levorato. Starting from the initial graph, the algorithm removes the vertex with highest degree until the graph is acyclic.
\end{description}

This class also provides a method that enumerates all the cycles. This can be usefull to find all the cycles that are impacted by a given vertex of the minimum feedback vertex set.

\begin{description}
\item[Set<ArrayList<V> > enumAllCircuitsTarjan(Graph<V,E> G)] Enumerates all circuits of a graph (Tarjan, \emph{Enumeration of the elementary
 circuits of a directed graph}~\cite{Tarjan1972}).
\end{description}

\subsubsection{Class MIS}

This class computes the maximum independ set of a graph (directed or undirected). This problem consists in finding the maximum set of vertices such that two vertices of the computed set are not neighbors in the original graph. Several algorithms have been implemented.

\begin{description}
\item[Set<V> maximalIndependentSetGreedy(Graph<V,E> g):] implements a polynomial greedy heuristic in $O(n+m)$. The algorithm chooses vertices of minimum degree and removes the neighbors iteratively.
\item[Set<V> maximumIndependentSetBruteForce(Graph<V,E> g):] this algorithm examines every vertex subset and checks whether it is an independent set using the \jv{isIndependentSet} method. The time complexity is $O(n^2.2^n)$.
\item[Set<V> maximumIndependentSetMaximumDegree(Graph<V,E> g):] this algorithm computes an exact solution by branching on the maximum degree vertex that is either in or out of the final solution and computes the result recursively. This algorithm has been proposed by I. Todinca and M. Liedloff but has no proved upper bound.
\item[Set<V> maximumIndependentSetMoonMoser(Graph<V,E> g):] the algorithm solves the problem in time $O(1.4423^n)$. It is based on \emph{On cliques in graphs}~\cite{Moon1965}.
\item[Set<V> maximumIndependentSetMoonMoserNonRecursive(Graph<V,E> g):] this is the non recurvisve version of the previous algorithm. Experimental benchmarks show that this version is slower.
\item[Set<V> maximuRmIndependentSetFominGrandoniKratsch(Graph<V,E> g):] this method is based on \emph{A Measure \& Conquer Approach for the Analysis of Exact Algorithms}~\cite{Fomin2009}. The algorithm solves the problem in time $O(1.2201^n)$.
\end{description}

Two methods help to detect if a set is an independent set for a given graph:

\begin{description}
\item[boolean isIndependentSet(Graph<V,E> g, Set<V> S):] verifies that S is an independent set of G.
\item[boolean isMaximalIndependentSet(Graph<V,E> g, Set<V> S):] verifies that S is a maximal independent set of G (cannot be completed).
\end{description}

\subsubsection{Class MVC}

This class implements algorithms for solving the Minimum Vertex Cover problem. The problem consists in finding the minimum set of vertices such that any vertex is at distance at most 1 of the computed set. This class only works on \textbf{undirected graphs}.

\begin{description}
\item[Set<V> twoApproximationCover(Graph<V,E> g):] 2-approximation of the Minimum Vertex Cover problem.  The algorithm returns a cover that is at most of double size of a minimal cover,  $O(|E|)$ time.
\item[Set<V> greedyCoverMaxDegree(Graph<V,E> g):] this greedy heuristic computes a result by selecting iteratively the vertex having the maximum  degree.
\item[boolean kVertexCoverBruteForce(Graph<V,E> g, int k):] this method selects all the sets of size $k$ among $n$ and checks if this set covers the graph. The resulting complexity is $O( {n \choose k})$. The algorithm tries to potentially select the vertices that are not covered first which improves the execution time.
\item[boolean kVertexCoverDegreeBranchingStrategy(Graph<V,E> g, int k):] this method uses Degree-Branching-Strategy (DBS) and has a time complexity of $O(1.47^k)$. This algorithm is extracted from \emph{Invitation to Fixed-Parameter Algorithms}~\cite{Niedermeier2006a} pp. 90.
\item[boolean kVertexCoverNiedermeier(Graph<V,E> g, int k):] this method is an implementation of Niedermeier algorithm of \emph{Invitation to Fixed Parameter Algorithms}~\cite{Niedermeier2006a} pp. 99-101, which time complexity is $O(1.33^k)$.
\item[boolean kVertexCoverBussGoldsmith(Graph<V,E> g, int k):] this method is an implementation of Buss and Goldsmith reduction algorithm of \emph{Nondeterminism within P}~\cite{Buss1993}, presented by Niedermeier in~\cite{Niedermeier2006a} pp. 54.

\end{description}

\subsubsection{Class Separators}

This class implements algorithms for solving the minimal separator problem. The AB-separator problem consists in finding the minimum set of vertices such that removing this set of vertices disconnects the vertex A from B.

\begin{description}
 \item[Set<Set<V> > getABSeparators(Graph<V,E> g, V a, V b):] this method computes the set of minimal ab-separator in $O(n^3)$ per separator. It is an implementation based on \emph{Generating All the Minimal Separators of a Graph}~\cite{Berry1999}.
 \item[Set<Set<V> > getAllMinimalSeparators(Graph<V,E> g):] this method returns all the minimal ab-separators for all pairs of vertices.
\end{description}

\subsection{Package agape.applications}  

This package contain two software: a graph graphical editor and command line program that allows to launch the previously presented algorithms. 

\subsubsection{Graphical editor: the GraphEditor class}

\begin{figure}[tb]
\begin{center}
\includegraphics[width=3.5in]{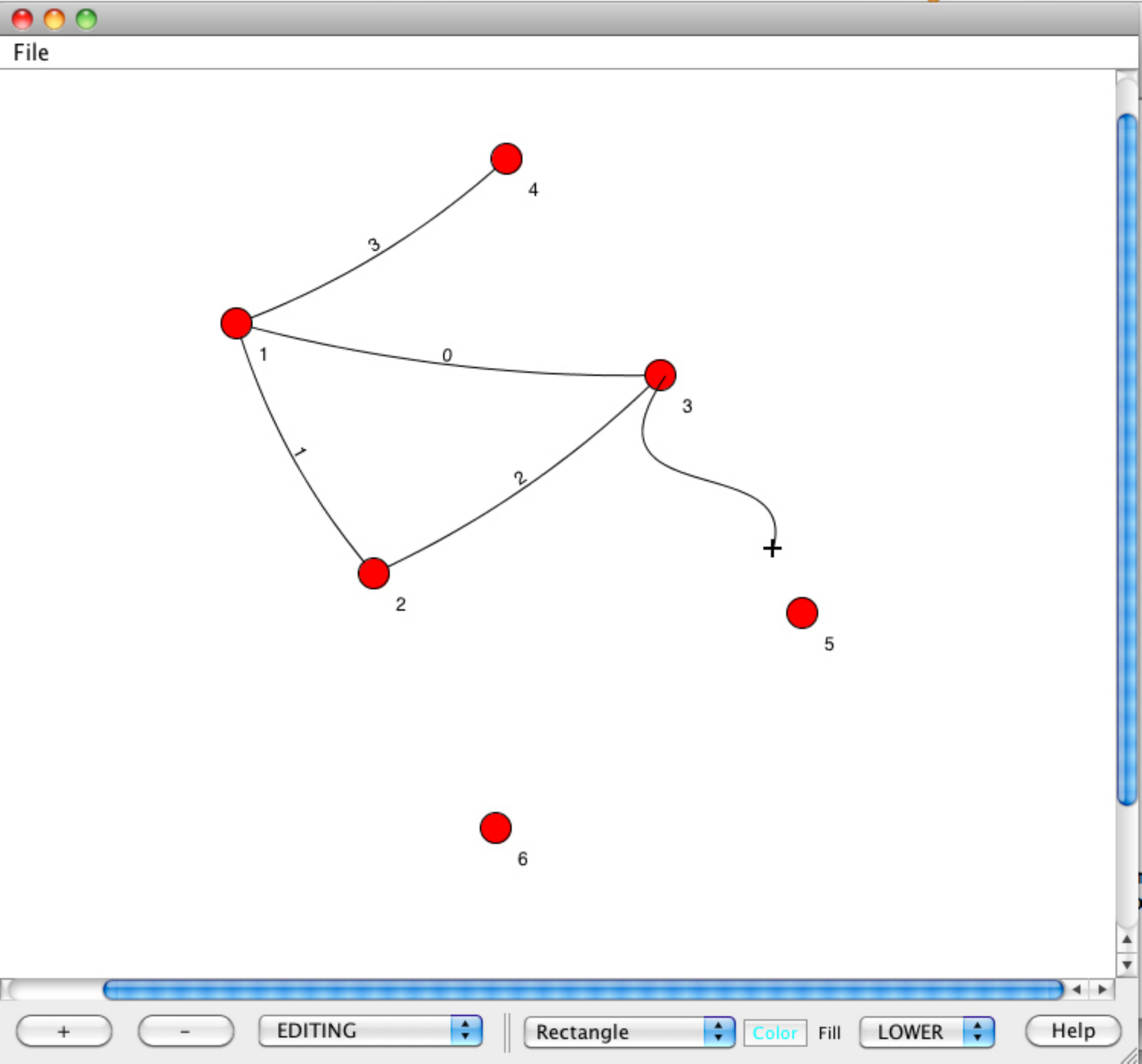}
\caption{Screen capture of the graph editor}
\label{fig:edit}
\end{center}
\end{figure}

The graphical editor is a simple example, based on the Jung library, as shown in Figure~\ref{fig:edit}. Only the possibly of saving a graph into a file has been added.

\subsubsection{Agape command line application: the AgapeCL class}
\label{sec:CL}

The command line application allows to apply an chosen algorithm to a graph or a set of graphs. All the graphs must be written in the \verb+.net+ format. This format comes from the Pajek software for social network analysis. A \verb+.net+ file is composed of the list of vertices and the adjacency list. An example of a Pajek file is presented in Listing~\ref{ex:net}. The graph is composed of 4 vertices and 4 edges (a,b), (b,c), (b,d), and (c,d).

\begin{figure}
\begin{lstlisting}[style=code, label=ex:net, title=Example of .net file]
*Vertices 4
1 "a"
2 "b"
3 "c"
4 "d"
*edgeslist
1 2 
2 3 4
3 4
\end{lstlisting}
\end{figure}

The \jv{AgapeCL} command line application can be used with the standalone package distributed on the website:

\begin{lstlisting}[style=bash]
java -jar AgapeCL.jar graphFilePath|graphDirectoryPath algorithm
\end{lstlisting}

The first parameter is the name of the \verb+.net+ file (or the directory that contains multiple files) and the second one is the algorithm name that has to be chosen among CN, MISBF, MISMM, MISDegMax, MISFGK, MVCBF, MVCBG, MVCDBS, MVCN, DFVS, SEP that correspond to the algorithm of Table~\ref{tab:algos}. The two following listings show examples of execution.

\begin{multicols}{2}
\begin{lstlisting}[style=bash, label=ex:ex1, title=Example of execution on test.net]
java -jar AgapeCL.jar ../test.net MISFGK

V:50 E:238
660.0 ms
Size: 14
[v0, v48, v17, v18, v36, v6, v25, v13, v9, v40, v34, v44, v31, v32]
\end{lstlisting}
\vfill
\columnbreak
\begin{lstlisting}[style=bash, label=ex:ex2, title=Example of execution on a directory]
java -jar AgapeCL.jar ../GenGraphs/temp/ MISFGK

16cell.net
V:8 E:24
36.0 ms
Size: 2
[3, 2]

snark.net
V:12 E:15
5.0 ms
Size: 6
[2, 10, 0, 5, 8, 11]

thomassen20.net
V:20 E:37
18.0 ms
Size: 8
[1, 6, 19, 4, 17, 9, 14, 11]
\end{lstlisting}
\end{multicols}

\subsection{Package agape.generators}  

This package contains several graph generators. Two categories of generators have been implemented: random and non random generators. Some generators are based on Jung generators and others have been implemented from scratch. Below is listed all the available generators. All the generators need factories to instantiate vertices and edges.

\subsubsection{Random graph generators}

\begin{itemize}
 \item Erd\"os Rényi random graphs~\cite{Erdos1959}.
 \item Eppstein random graphs (from Jung)~\cite{Eppstein2002}.
 \item Barabasi-Albert random graphs (from Jung)~\cite{Barabasi1999}.
 \item Kleinberg small world graphs~\cite{Kleinberg2000}.
 \item Watts-Strogatz small world graphs~\cite{Watts1998}.
 \item k-regular random graphs.
\end{itemize}

\subsubsection{Structured graph generators}

See Section~\ref{tuto:generators} for an example of use of these two generators.

\begin{itemize}
 \item 2D grids.
 \item K-regular rings.
\end{itemize}

In order to use the generators, the user has to define first a graph factory before using one of the generators. A tutorial about factories is presented in Section~\ref{tuto:factories}.
 
\subsection{Package agape.io}

This package contains two classes for the read/write operation on graphs (\jv{Import} and \jv{Export}).
 
Formats supported for reading:

\begin{itemize}
\item \verb+.net+ oriented or non oriented (Pajek)~\cite{DeNooy2005}.
\item \verb+.mgl+ Mascopt~\cite{LALANDE:2004:INRIA-00069887:1}.
\item \verb+.tgf+ Wolfram Mathematica \url{http://www.wolfram.com/mathematica/}.
\end{itemize}

Formats supported for writting:

\begin{itemize}
\item \verb+.net+ oriented or non oriented (Pajek)~\cite{DeNooy2005}.
\item \verb+.gv+  GraphViz \url{http://www.graphviz.org/}.
\end{itemize}

A tutorial about input/output is presented in section~\ref{tuto:io}.

\subsection{Package agape.tools}

This package is a set of toolboxes for classical graph operations. In the \jv{Operations} class, we implemented metric detection (diameter, degrees, \ldots), type identification (clique, regular, simple edge, \ldots), copy operations (copy, merge, subgraphs, \ldots), \ldots For example, the \jv{getMinDeg(Graph<V,E> g)} returns the vertex that has the smallest degree and \jv{isRegular(Graph<V,E> g} test if the graph is k-regular. In the \jv{Components} class, we implemented methods for connected components of a graph, for example the Tarjan's method that computes all the strongly connected components in \jv{getAllStronglyConnectedComponent(Graph<V,E> g}.

\subsection{Package agape.visualization}  

\begin{figure}[tb]
\begin{center}
\includegraphics[width=3in]{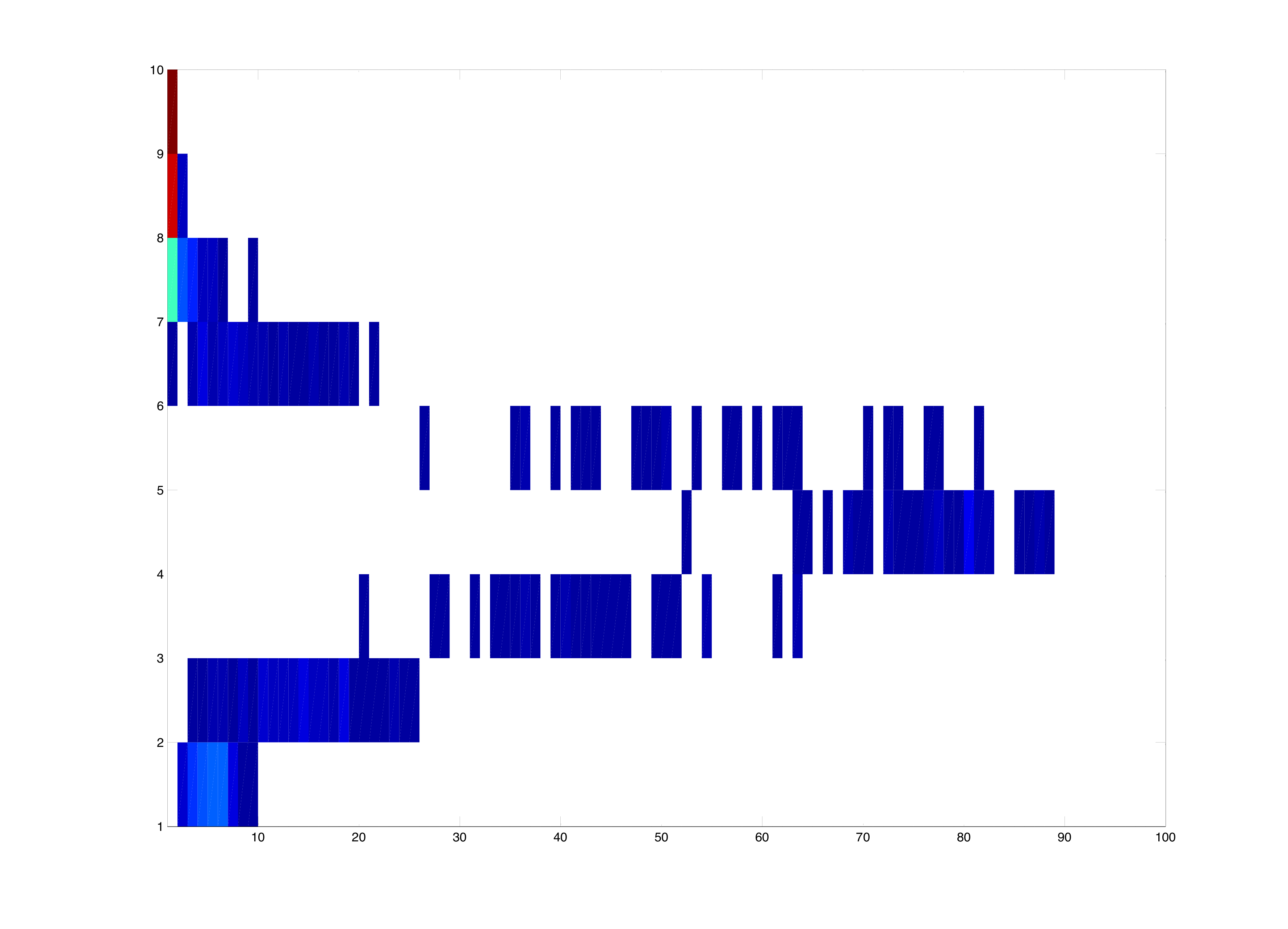}
\end{center}
\caption{B-matrix for a Watts-STrogatz small world}
\label{fig:bmatrix}
\end{figure}

This package contains the class Visualization that provides a method to display a graph (\jv{showGraph}) and a method to represent a graph as a B-matrix~\cite{Bagrow2007}, as shown in Figure~\ref{fig:bmatrix}.
 
\newpage

\section{Tutorials}
\label{sec:tutorials}
\subsection{Tutorials for basics}

\subsubsection{Creating Jung graphs}

Jung graphs are easy to build using templated classes such as \jv{DirectedSparseGraph}. Vertices and edges are java objects that are manipulated by the Jung class graphs.

\begin{multicols}{2}
\begin{lstlisting}[style=java,title=Tutorial: agape.tutorials.JungGraphTutorial]
/*
 * Copyright University of Orleans - ENSI de Bourges. 
 * Source code under CeCILL license.
 */
package agape.tutorials;

import edu.uci.ics.jung.graph.DirectedSparseGraph;
import edu.uci.ics.jung.graph.util.Pair;

public class JungGraphTutorial {
    public static void main(String[] args) {

        DirectedSparseGraph<String, Integer> g = new DirectedSparseGraph<String, Integer>();
        g.addVertex("n1");
        g.addVertex("n2");
        g.addVertex("n3");
        // Jung finds matching nodes even if  pointers are different
        g.addEdge(1, new Pair<String>("n1", "n2"));
        System.out.println(g);
        // Jung adds automatically new nodes
        g.addEdge(2, new Pair<String>("n1", "n4"));
        System.out.println(g);
        g.removeVertex("n1");
        System.out.println(g);
    }
}
\end{lstlisting}
\columnbreak
\begin{lstlisting}[style=java,title=Tutorial: console output]
Vertices:n1,n3,n2
Edges:1[n1,n2] 
Vertices:n1,n4,n3,n2
Edges:1[n1,n2] 2[n1,n4] 
Vertices:n4,n3,n2
Edges:
\end{lstlisting}
\end{multicols}

\newpage
\subsubsection{Input/Output}
\label{tuto:io}

A .net reader have been implemented in the class \jv{Import}. The first line of the file is ignored (title), then each line is parsed to read vertices (name ``label'') until the string \jv{*edgeslist} is found. Then, the edges/arcs are read until the end. The .net file instanciates \jv{Graph<String,Integer>} which implies that vertices are \jv{String} and edges/arcs are \jv{Integer}. Exporting a graph to a file works the same way. The \jv{Export} class can also write files using the graphviz format.

\begin{multicols}{2}
\begin{lstlisting}[style=java,title=Tutorial: agape.tutorials.IOTutorial]
/*
 * Copyright University of Orleans - ENSI de Bourges. 
 * Source code under CeCILL license.
 */
package agape.tutorials;

import java.io.IOException;

import agape.io.Export;
import agape.io.Import;
import edu.uci.ics.jung.graph.Graph;

/**
 * @author jf
 */
public class IOTutorial {
	public static void main(String[] args) {
		Graph<String,Integer> g = Import.readDNet("src/agape/tutorials/IOTutorial.net");
		System.out.println(g);
		try {
			Export.writeGV("src/agape/tutorials/IOTutorial", g);
		} catch (IOException e) {		
			e.printStackTrace();
		}
	}
}
\end{lstlisting}
\columnbreak
\begin{lstlisting}[style=java,title=Tutorial: IOTutorial.net]
*Vertices 4
1 "a"
2 "b"
3 "c"
4 "d"
*edgeslist
1 2
2 3 4
3 4
\end{lstlisting}
\begin{lstlisting}[style=java,title=Tutorial: IOTutorial.gv]
digraph G {
size="100,20"; ratio = auto;
node [style=filled];
"a" -> "b" [color="0.649 0.701 0.701"];
"b" -> "c" [color="0.649 0.701 0.701"];
"b" -> "d" [color="0.649 0.701 0.701"];
"c" -> "d" [color="0.649 0.701 0.701"];
"d" [color=grey];
"b" [color=grey];
"c" [color=grey];
"a" [color=grey];
}
\end{lstlisting}
\begin{center}
\includegraphics[width=\columnwidth]{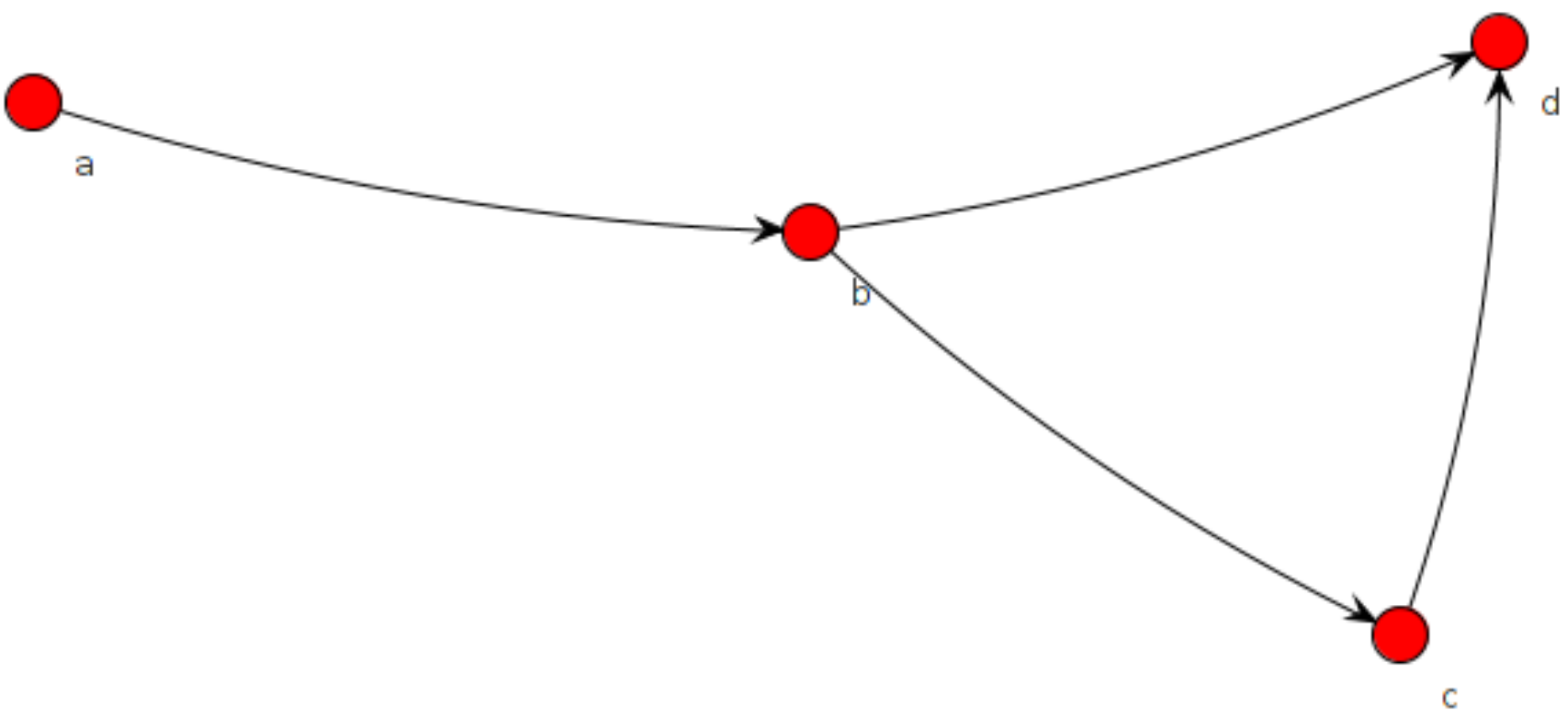}
\end{center}
\end{multicols}

\newpage
\subsubsection{Using factories}
\label{tuto:factories}

Several algorithms, especially the developed generators use factories to create new graphs. Indeed, the algorithm does not know the real type of the graph that implements the Jung interface. For example, a graph that has vertices based on \jv{String} and edges on \jv{Integer}, is typed as \jv{Graph<String, Integer>}. The definition of factories is based on the interface \jv{Factory} of apache's commons that needs to be instanciated with the choosen type: in order to build a \jv{new MyFactory<String>()}, \jv{MyFactory} must implement the \jv{create()} method.

\begin{multicols}{2}
\begin{lstlisting}[style=java,title=Tutorial: agape.tutorials.UsingFactoriesTutorial]
/*
 * Copyright University of Orleans - ENSI de Bourges. 
 * Source code under CeCILL license.
 */
package agape.tutorials;

import org.apache.commons.collections15.Factory;

import agape.generators.RandGenerator;
import agape.visu.Visualization;
import edu.uci.ics.jung.graph.Graph;
import edu.uci.ics.jung.graph.UndirectedSparseGraph;

public class UsingFactoriesTutorial {
	public static void main(String[] args) {

		Factory<Integer> edgeFactory = new Factory<Integer>() {
			int c=0;
			public Integer create() {
				c++;
				return Integer.valueOf(c);
			}
		};

		Factory<String> vertexFactory = new Factory<String>() {
			int c=0;
			public String create() {
				c++;
				return "v"+c;
			}
		};

		Factory<Graph<String,Integer>> graphFactory = new Factory<Graph<String,Integer>>() {
			public Graph<String,Integer> create() {
				return new UndirectedSparseGraph<String, Integer>();
			}
		};

		RandGenerator<String, Integer> generator = new RandGenerator<String, Integer>();
		Graph<String, Integer> g = generator.generateErdosRenyiGraph(graphFactory, vertexFactory, edgeFactory, 10, 3);
		System.out.println(g);

		Visualization.showGraph(g);
	}
}
\end{lstlisting}
\columnbreak
\begin{lstlisting}[style=java,title=Tutorial: console output]
Vertices:v1,v7,v5,v6,v4,v9,v3,v8,v2,v10
Edges:1[v1,v7] 2[v1,v5] 3[v1,v6] 4[v1,v4] 5[v1,v9] 6[v1,v3] 7[v1,v8] 8[v1,v2] 9[v1,v10] 10[v7,v5] 11[v7,v6] 12[v7,v4] 13[v7,v9] 14[v7,v3] 15[v7,v8] 17[v7,v10] 16[v7,v2] 19[v5,v4] 18[v5,v6] 21[v5,v3] 20[v5,v9] 23[v5,v2] 22[v5,v8] 25[v6,v4] 24[v5,v10] 27[v6,v3] 26[v6,v9] 29[v6,v2] 28[v6,v8] 31[v4,v9] 30[v6,v10] 34[v4,v2] 35[v4,v10] 32[v4,v3] 33[v4,v8] 38[v9,v2] 39[v9,v10] 36[v9,v3] 37[v9,v8] 42[v3,v10] 43[v8,v2] 40[v3,v8] 41[v3,v2] 44[v8,v10] 45[v2,v10] 
\end{lstlisting}

\begin{center}
\includegraphics[width=0.7\columnwidth]{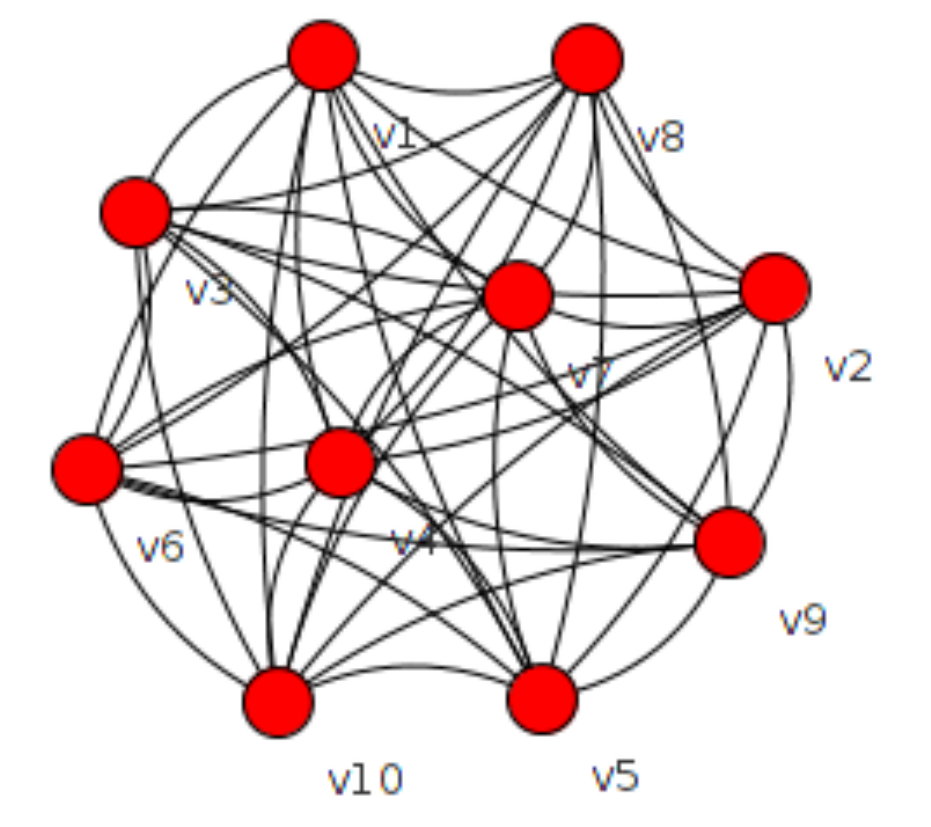}
\end{center}

\end{multicols}

\newpage

As a consequence, some operations have been implemented using factories. For example, the \jv{copyGraph} method asks for a factory in order to copy any kind of graph, as shown in the example below. Furthermore, many algorithms need to manipulate deeply graphs by creating, deleting edges and/or vertices. These algorithms thus require the adequate factories to complete.

\begin{multicols}{2}
\begin{lstlisting}[style=java,title=Tutorial: agape.tutorials.CopyGraphTutorial]
/*
 * Copyright University of Orleans - ENSI de Bourges. 
 * Source code under CeCILL license.
 */
package agape.tutorials;

import agape.tools.Operations;
import edu.uci.ics.jung.graph.SparseGraph;
import edu.uci.ics.jung.graph.UndirectedSparseGraph;
import edu.uci.ics.jung.graph.util.Pair;

public class CopyGraphTutorial {

    public static void main(String[] args) {
        System.out.println("----------------");
        System.out.println("UNDIRECTED GRAPH");
        System.out.println("----------------");
        SparseGraph<String, Integer> gu = new SparseGraph<String, Integer>();
        UndirectedGraphFactoryForStringInteger undfactory = new UndirectedGraphFactoryForStringInteger();
        
        gu.addVertex("n1");  gu.addVertex("n2");
        gu.addVertex("n3");  gu.addVertex("n4");
        gu.addVertex("n5");
        
        gu.addEdge(1, new Pair<String>("n1", "n2"));
        gu.addEdge(2, new Pair<String>("n1", "n4"));
        gu.addEdge(3, new Pair<String>("n2", "n3"));
        gu.addEdge(4, new Pair<String>("n3", "n5"));
        gu.addEdge(5, new Pair<String>("n5", "n2"));
        gu.addEdge(6, new Pair<String>("n5", "n3")); // useless
        
        /* This is an explicit call to the copyUndirectedSparseGraph method
         * (we know that the graph is undirected)
         */
        System.out.println(gu);
        UndirectedSparseGraph<String, Integer> gu2 = 
        		(UndirectedSparseGraph<String, Integer>) Operations.copyUndirectedSparseGraph(gu);
        System.out.println(gu2);
        /* This is a general call to the copyGraph method that uses a factory
         * (we do not want to know if the graph is direted or not)
         */
        UndirectedSparseGraph<String, Integer> gu3 = 
        		(UndirectedSparseGraph<String, Integer>) Operations.copyGraph(gu, undfactory);
        System.out.println(gu3);

    }
}
\end{lstlisting}
\columnbreak
~
\end{multicols}

\newpage

Using factories may cause issues when graphs are created manually because there may be conflicts between the generated vertices/edges and the already existing ones. The following example shows such a conflict. To solve the issue, the user should create any graph using the factory that is used later in the algorithms.

\begin{multicols}{2}
\begin{lstlisting}[style=java,title=Tutorial: agape.tutorials.FactoryProblem]
/*
 * Copyright University of Orleans - ENSI de Bourges. 
 * Source code under CeCILL license.
 */
package agape.tutorials;

import java.util.Iterator;

import org.apache.commons.collections15.Factory;

import edu.uci.ics.jung.graph.Graph;
import edu.uci.ics.jung.graph.SparseGraph;
import edu.uci.ics.jung.graph.util.Pair;

public class FactoryProblem<V,E> {

    /**
     * @param args
     */
    public static void main(String[] args) {
        
        // Creating the graph and the corresponding factory
        UndirectedGraphFactoryForStringInteger undfactory = new UndirectedGraphFactoryForStringInteger();
        
        SparseGraph<String, Integer> gu = new SparseGraph<String, Integer>();
        gu.addVertex("v1");  gu.addVertex("v2");
        gu.addVertex("v3");  gu.addVertex("v4");
        gu.addVertex("v5");
        
        gu.addEdge(1, new Pair<String>("v1", "v2"));
        gu.addEdge(2, new Pair<String>("v1", "v4"));
        gu.addEdge(3, new Pair<String>("v2", "v3"));
        gu.addEdge(4, new Pair<String>("v3", "v5"));
        gu.addEdge(5, new Pair<String>("v5", "v2"));
        gu.addEdge(6, new Pair<String>("v5", "v3"));
        
        FactoryProblem<String, Integer> problem = new FactoryProblem<String, Integer>();
        try
        {
           problem.addRandomEdge(gu, undfactory.edgeFactory);
        }
        catch (IllegalArgumentException e)
        {
            System.err.println(e);
            System.out.println("Our factory is badly used because it tries to instantiate the edge number 1 " +
            		"which already exists in the graph.");
        }
        
        // Solution: we should not have created the graph manually !
        SparseGraph<String, Integer> gu2 = new SparseGraph<String, Integer>();
        String v[] = new String[5];
        for (int i=0; i<5; i++)
        {
            v[i] = undfactory.vertexFactory.create();
            gu2.addVertex(v[i]);
        }
        Integer e[] = new Integer[6];
        for (int i=0; i<6; i++)
            e[i] = undfactory.edgeFactory.create();
        gu2.addEdge(e[0], v[0], v[1]);
        gu2.addEdge(e[1], v[0], v[3]);
        gu2.addEdge(e[2], v[1], v[2]);
        gu2.addEdge(e[3], v[2], v[4]);
        gu2.addEdge(e[4], v[4], v[2]);
        gu2.addEdge(e[5], v[4], v[3]);
        
        System.out.println("gu2 before : " + gu2);
        problem.addRandomEdge(gu2, undfactory.edgeFactory);
        System.out.println("gu2 after  : " + gu2);
    }
    
    public void addRandomEdge(Graph<V, E> g, Factory<E> f)
    {
        // Choosing 2 vertices
        int choice = (int)(Math.random()*g.getVertexCount()) ;
        Iterator<V> it = g.getVertices().iterator();
        V v1 = null;
        for (int i = 0; i <= choice; i++)
            v1 = it.next();
       
        choice = (int)(Math.random()*g.getVertexCount()) ;
        it = g.getVertices().iterator();
        V v2 = null;
        for (int i = 0; i <= choice; i++)
            v2 = it.next();
               
        E edge = f.create();
        System.out.println("Trying to add Edge " + edge + " between " + v1 + " " + v2);
        g.addEdge(edge, v1, v2);
    }
}
\end{lstlisting}
\begin{lstlisting}[style=java,title=Tutorial: console output]
Trying to add Edge 1 between v4 v3
java.lang.IllegalArgumentException: edge 1 already exists in this graph with endpoints <v1, v2> and cannot be added with endpoints <v4, v3>
Our factory is badly used because it tries to instantiate the edge number 1 which already exists in the graph.
gu2 before : Vertices:v1,v5,v4,v3,v2
Edges:2[v1,v2] 3[v1,v4] 4[v2,v3] 5[v3,v5] 7[v5,v4] 
Trying to add Edge 8 between v2 v4
gu2 after  : Vertices:v1,v5,v4,v3,v2
Edges:2[v1,v2] 3[v1,v4] 4[v2,v3] 5[v3,v5] 7[v5,v4] 8[v2,v4] 

\end{lstlisting}
\end{multicols}

\newpage
\subsubsection{Using generators}
\label{tuto:generators}

\begin{multicols}{2}
\begin{lstlisting}[style=java,title=Tutorial: agape.tutorials.NRandomGeneratorTutorial]
/*
 * Copyright University of Orleans - ENSI de Bourges. 
 * Source code under CeCILL license.
 */
package agape.tutorials;

import agape.generators.NRandGenerator;
import agape.visu.Visualization;
import edu.uci.ics.jung.algorithms.layout.CircleLayout;
import edu.uci.ics.jung.graph.Graph;

public class NRandomGeneratorTutorial {
	public static void main(String[] args) {

	    DirectedGraphFactoryForStringInteger factory = new DirectedGraphFactoryForStringInteger();
				
		Graph<String, Integer> g = NRandGenerator.generateRegularRing(factory, factory.vertexFactory, factory.edgeFactory, 20, 6);
		Visualization.showGraph(g, new CircleLayout<String,Integer>(g));
		
		Graph<String, Integer> g2 = NRandGenerator.generateGridGraph(factory, factory.vertexFactory, factory.edgeFactory, 3, 5, false);
		Visualization.showGraph(g2);
		
	}
}
\end{lstlisting}

\begin{center}
\includegraphics[width=0.8\columnwidth]{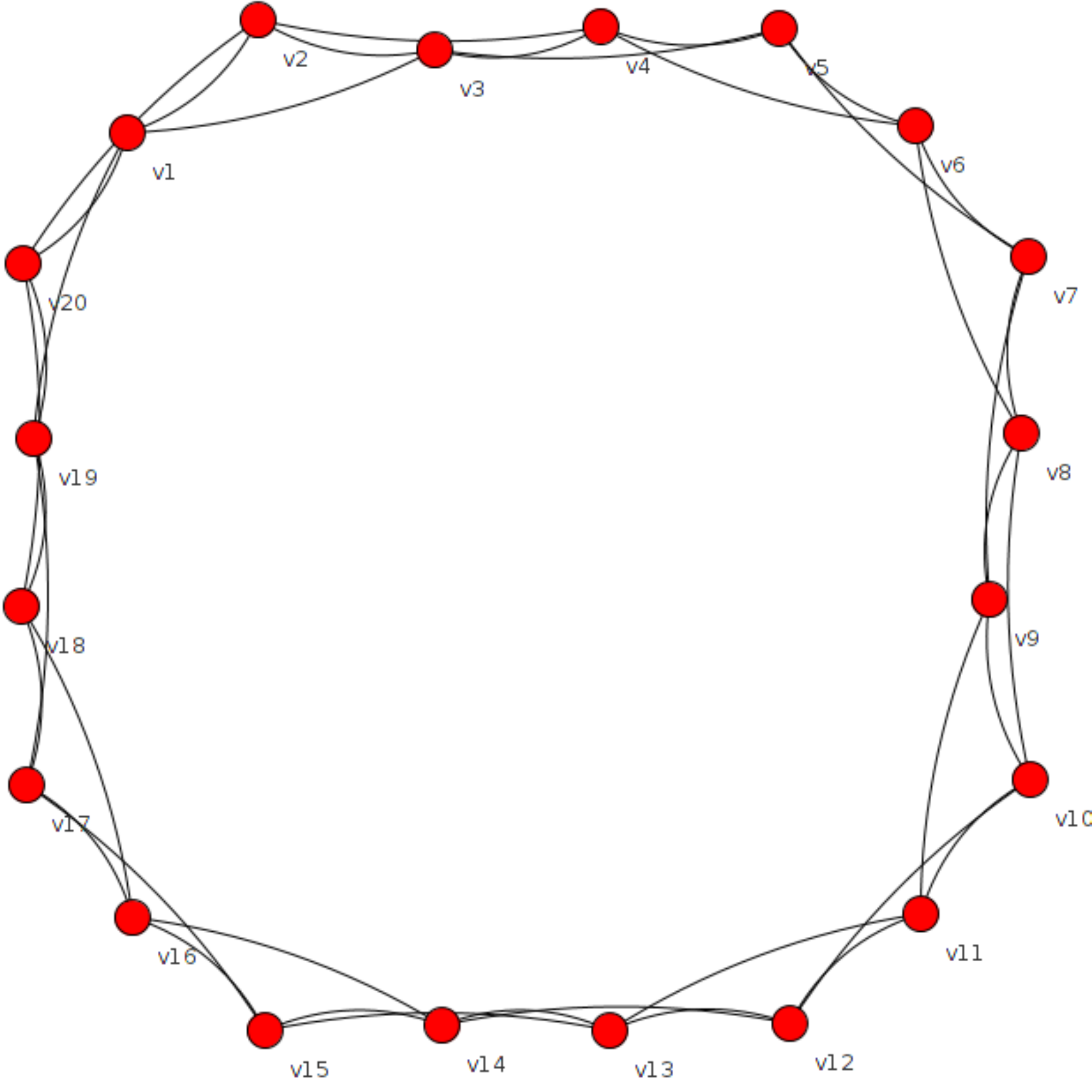}
\includegraphics[width=0.8\columnwidth]{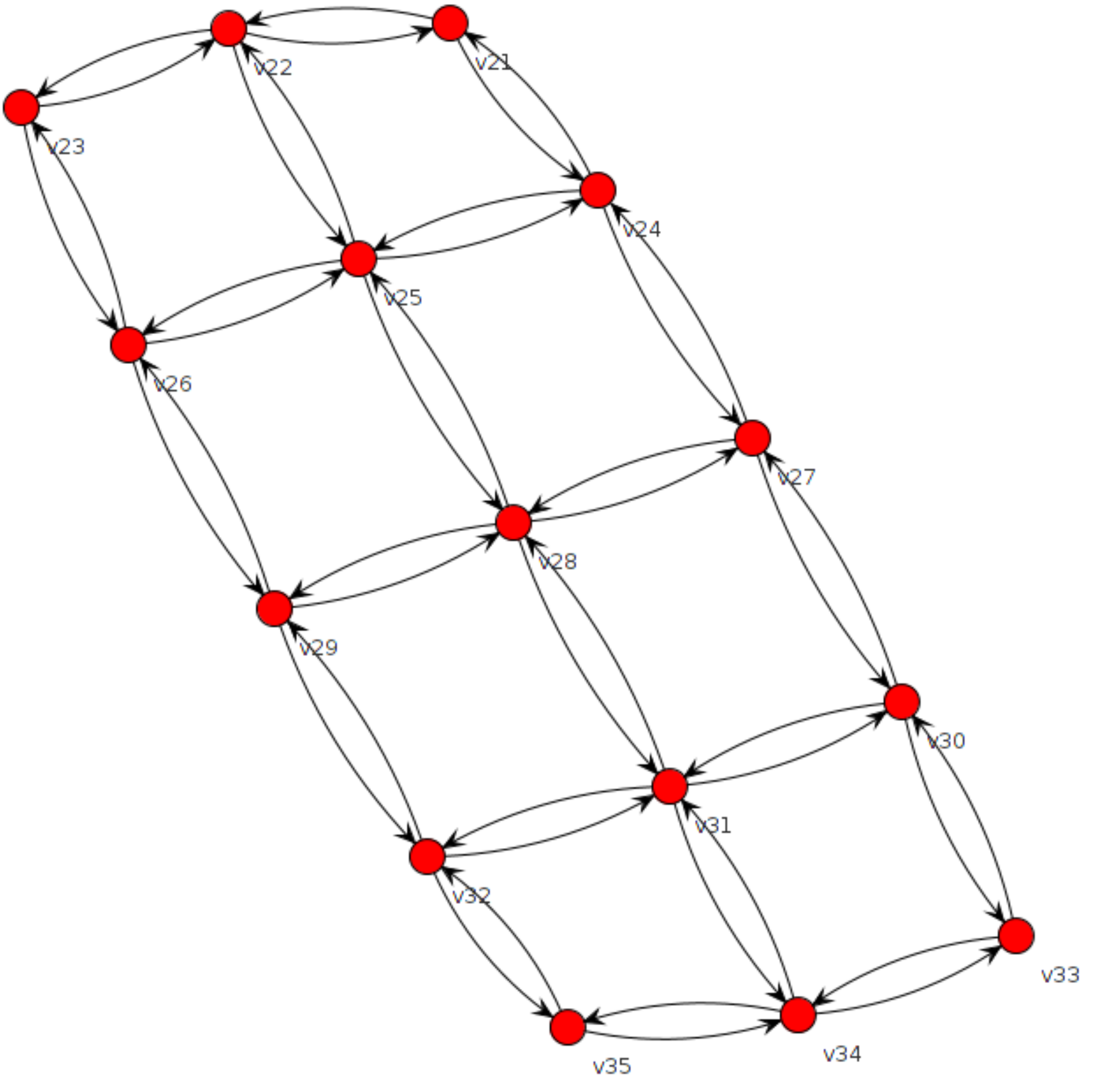}
\end{center}
\columnbreak

\begin{lstlisting}[style=java,title=Tutorial: agape.tutorials.RandomGeneratorTutorial]
/*
 * Copyright University of Orleans - ENSI de Bourges. 
 * Source code under CeCILL license.
 */
package agape.tutorials;

import agape.generators.RandGenerator;
import agape.visu.Visualization;
import edu.uci.ics.jung.algorithms.layout.CircleLayout;
import edu.uci.ics.jung.graph.Graph;

public class RandomGeneratorTutorial {
    public static void main(String[] args) throws Exception {

        UndirectedGraphFactoryForStringInteger factory = new UndirectedGraphFactoryForStringInteger();

        Graph<String, Integer> g2 = RandGenerator.generateKleinbergSWGraph(
                factory, factory.vertexFactory, factory.edgeFactory, 5, 5, 1, 2, 2);
        Visualization.showGraph(g2);

        Graph<String, Integer> g3 = RandGenerator.generateWattsStrogatzSWGraph(
                factory, factory.vertexFactory, factory.edgeFactory, 24, 4, 0.5);
        Visualization.showGraph(g3, new CircleLayout<String,Integer>(g3));

        Graph<String, Integer> g4 = RandGenerator.generateErdosRenyiGraph(
                factory, factory.vertexFactory, factory.edgeFactory, 10, 0.5);
        Visualization.showGraph(g4);

        Graph<String, Integer> g5 = RandGenerator.generateBarabasiAlbertGraph(
                factory, factory.vertexFactory, factory.edgeFactory, 10, 5, 7);
        Visualization.showGraph(g5);

        Graph<String, Integer> g6 = RandGenerator.generateEppsteinGraph(
                factory, factory.vertexFactory, factory.edgeFactory, 10, 25, 100);
        Visualization.showGraph(g6);

        Graph<String, Integer> g7 = RandGenerator.generateRandomRegularGraph(
                factory, factory.vertexFactory, factory.edgeFactory, 10, 5);
        Visualization.showGraph(g7);
    }
}
\end{lstlisting}

\end{multicols}

\begin{center}
\begin{figure}[p]
\subfigure[g2: Kleinberg]{
\includegraphics[width=0.45\textwidth]{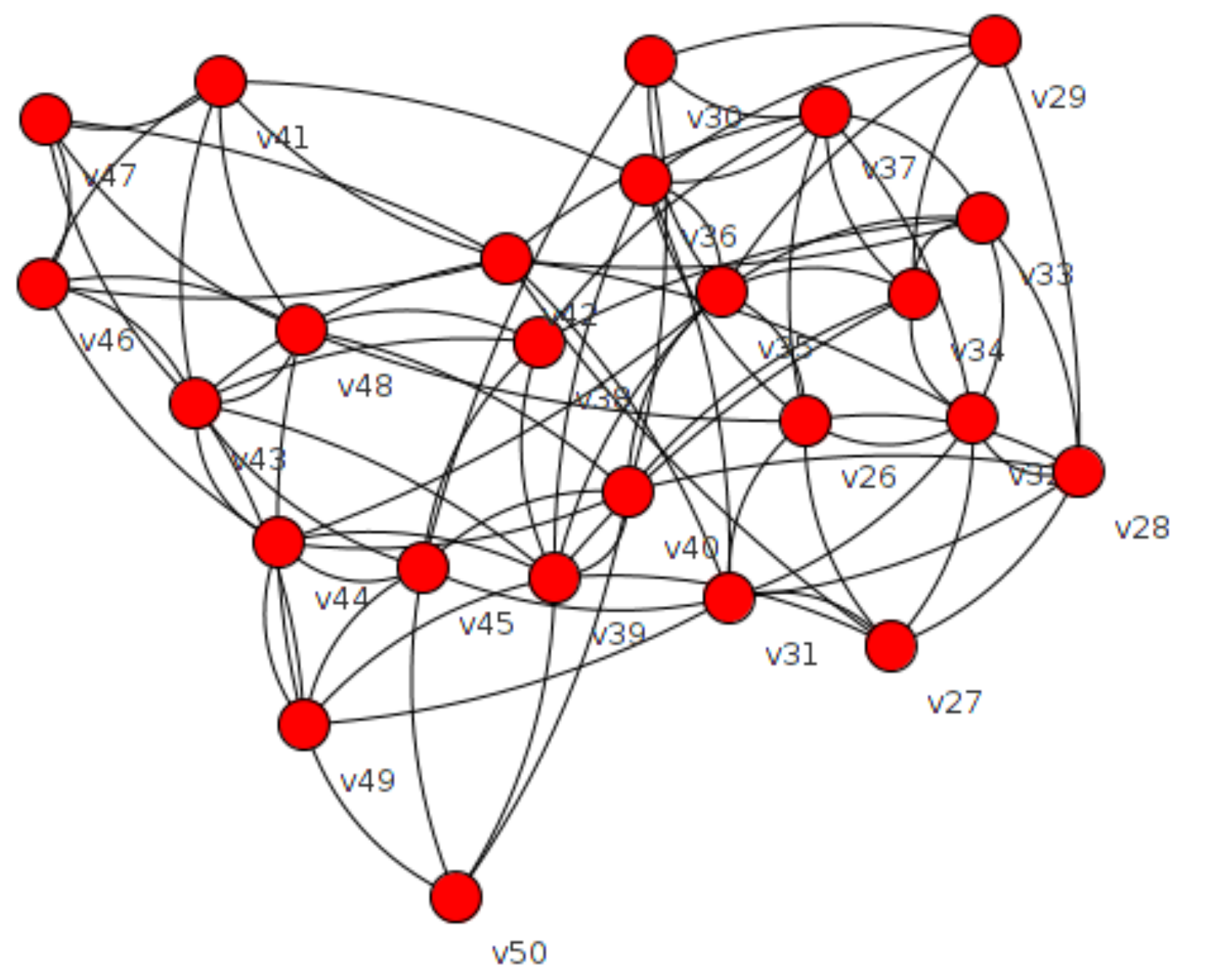}
}
\subfigure[g3: Watts-Strogatz small world graph]{
\includegraphics[width=0.45\textwidth]{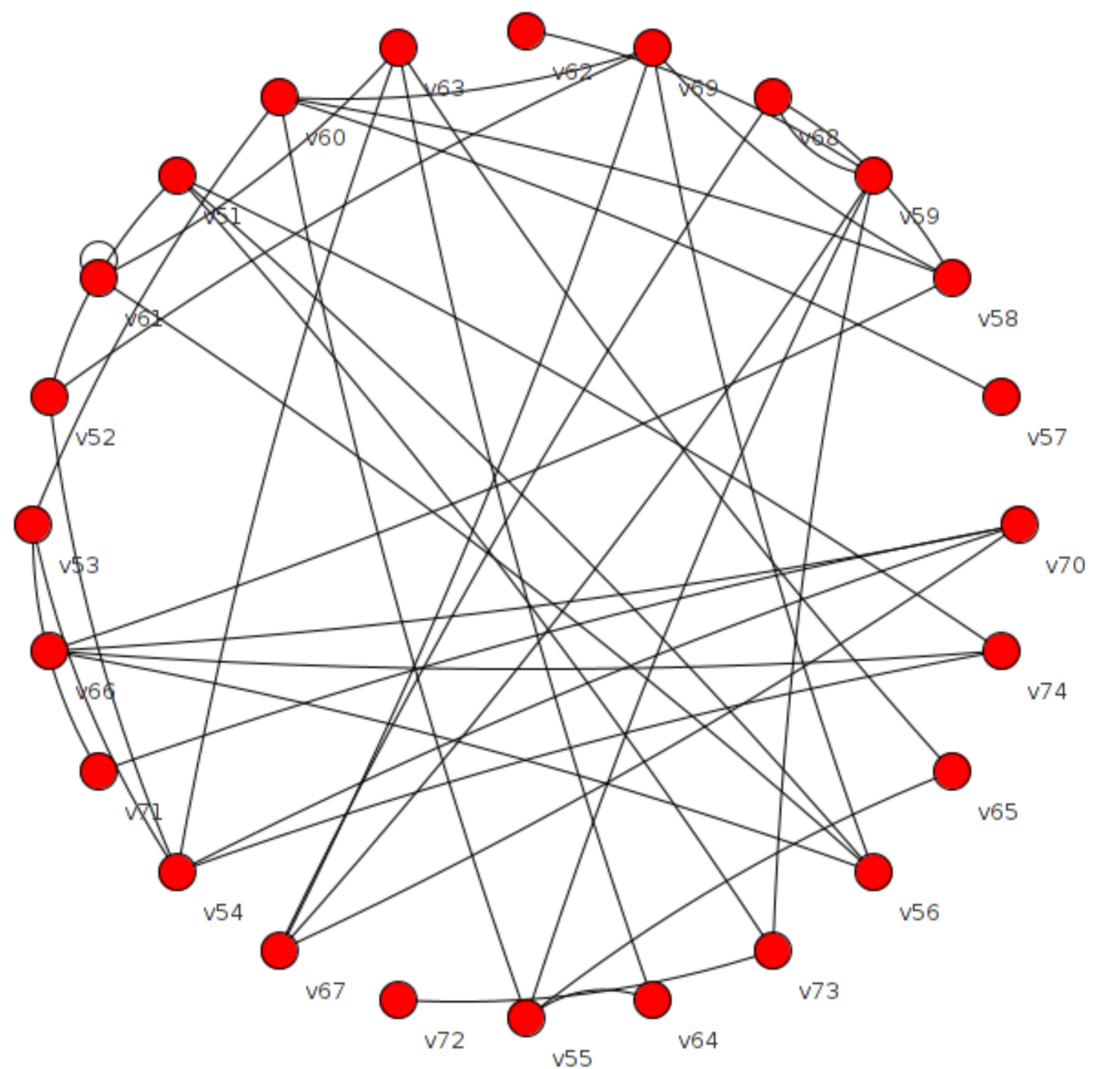}
}
\subfigure[g4: Erd\"os Rényi random graph]{
\includegraphics[width=0.40\textwidth]{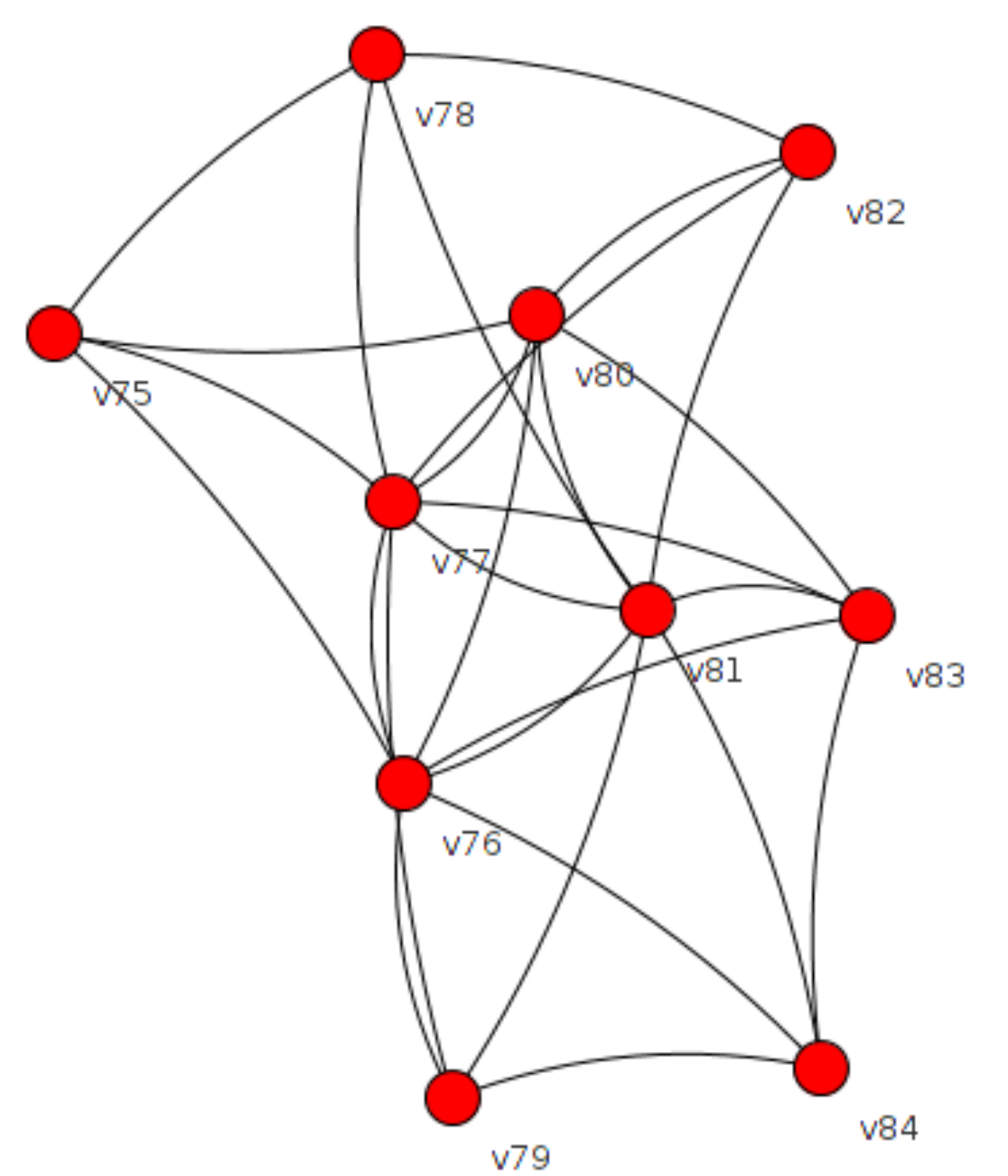}
}
\subfigure[g5 Barabasi-Albert random graph]{
\includegraphics[width=0.45\textwidth]{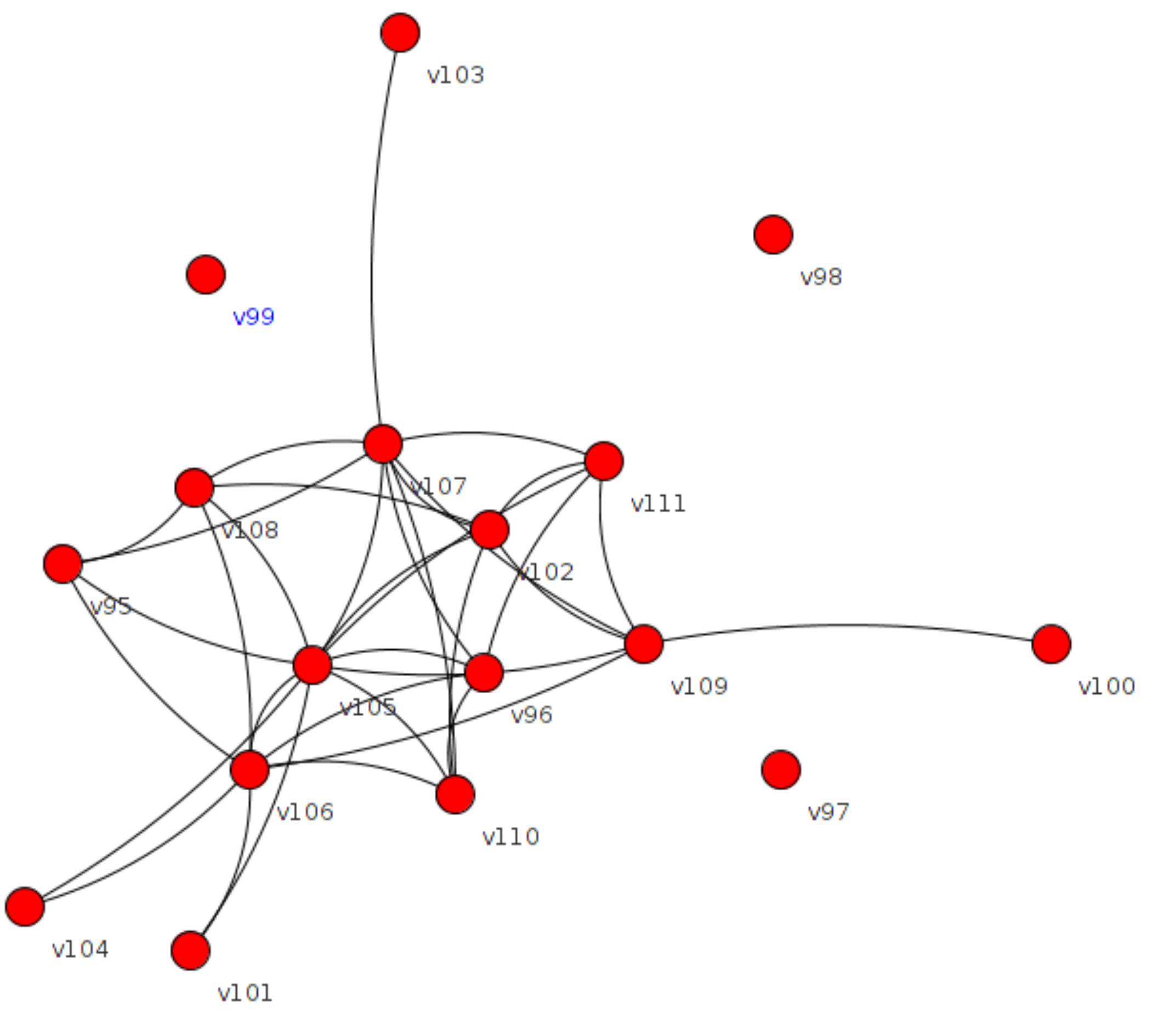}
}
\subfigure[g6: Eppstein random graph]{
\includegraphics[width=0.45\textwidth]{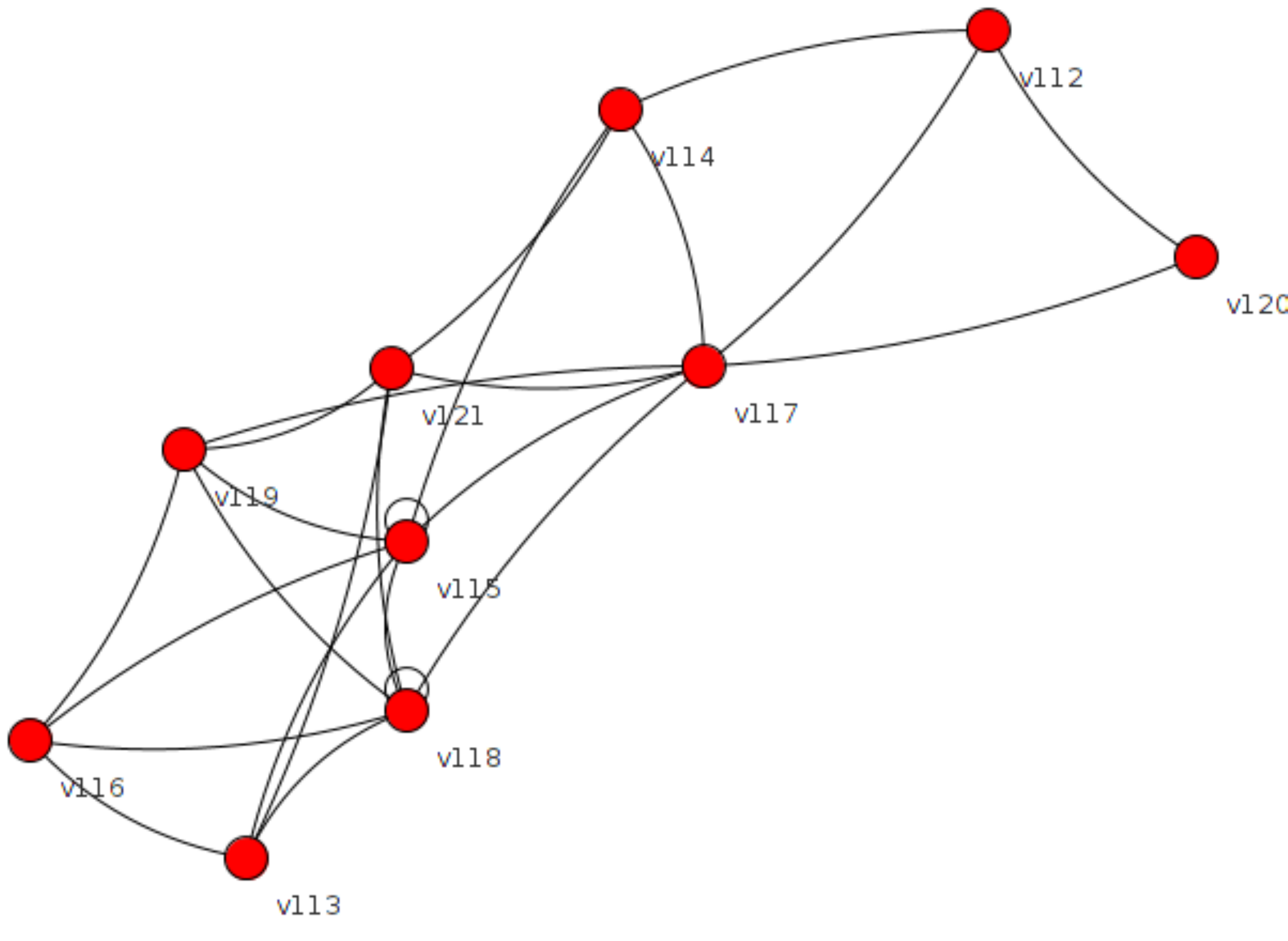}
}
\subfigure[g7: k-regular random graph]{
\includegraphics[width=0.45\textwidth]{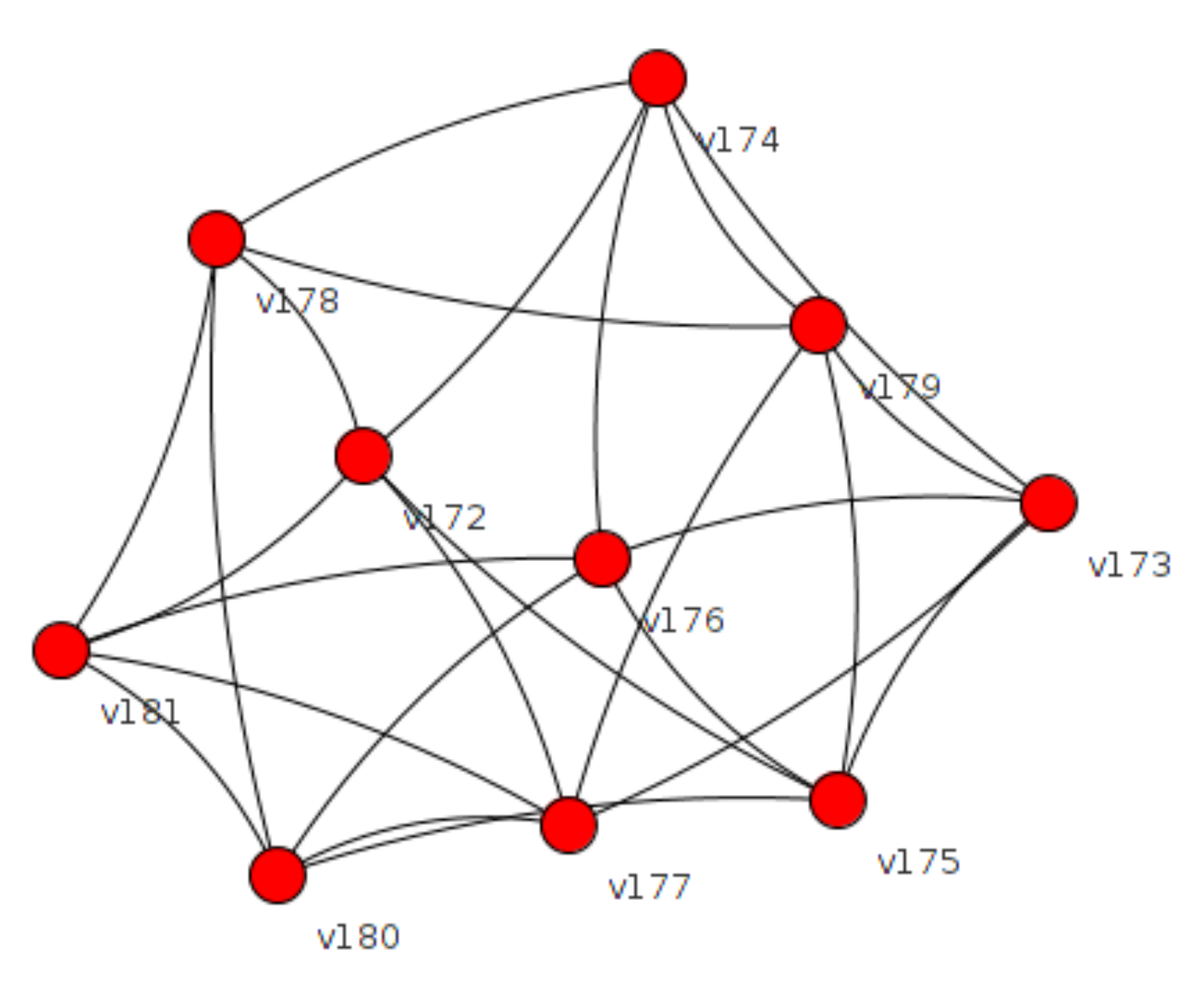}
}
\caption{RandGenerator outputs}
\end{figure}
\end{center}

\newpage
\subsection{Tutorials for algorithms}

\subsubsection{Using Coloring algorithms}
\label{tuto:coloring}

After generating a graph using one of the factories, this example shows how to compute the chromatic number of a graph using the \jv{Coloring} class.

\begin{multicols}{2}
\begin{lstlisting}[style=java,title=Tutorial: agape.algos.Coloring]
/*
 * Copyright University of Orleans - ENSI de Bourges. 
 * Source code under CeCILL license.
 */
package agape.tutorials;

import java.util.Set;

import agape.algos.Coloring;
import agape.generators.RandGenerator;
import agape.visu.Visualization;
import edu.uci.ics.jung.graph.Graph;

public class AlgoColoringTutorial {
	public static void main(String[] args) {

	    UndirectedGraphFactoryForStringInteger factory = new UndirectedGraphFactoryForStringInteger(1,1);
		RandGenerator<String, Integer> generator = new RandGenerator<String, Integer>();
		Graph<String, Integer> g = generator.generateErdosRenyiGraph(factory, factory.vertexFactory, factory.edgeFactory, 9, 0.7);
		System.out.println("Generating an Erdos Renyi graph:");
		System.out.println(g);
       
			
		Coloring<String,Integer> coloring = new Coloring<String, Integer>(factory);
		int col = coloring.chromaticNumber(g);
		System.out.println("Chromatic number: " + col);
        
		Set<Set<String>> sol = coloring.graphColoring(g);
		System.out.println(sol);
		
		Visualization.showGraphSets(g, sol);
		
		
		Graph<String, Integer> g2 = generator.generateErdosRenyiGraph(factory, factory.vertexFactory, factory.edgeFactory, 90, 0.05);
		System.out.println("Graph2: " + g2.getVertexCount() + " vertices, " + g2.getEdgeCount() + " edges.");
		Set<Set<String>> sol2 = coloring.greedyGraphColoring(g2);
		System.out.println("Greedy chromatic number: " + sol2.size());
		
		Visualization.showGraphSets(g2, sol2);
	}
}
\end{lstlisting}
\begin{lstlisting}[style=java,title=Tutorial: console output]
Generating an Erdos Renyi graph:
Vertices:v1,v7,v5,v6,v4,v9,v3,v8,v2
Edges:1[v1,v9] 2[v1,v3] 3[v1,v8] 4[v1,v2] 5[v7,v5] 6[v7,v6] 7[v7,v8] 8[v7,v2]
9[v5,v4] 10[v5,v9] 11[v5,v3] 12[v5,v8] 13[v5,v2] 14[v6,v9] 15[v6,v3] 17[v6,v2]
16[v6,v8] 19[v4,v8] 18[v4,v3] 21[v9,v3] 20[v4,v2] 23[v3,v2] 22[v3,v8] 
Chromatic number: 4
[[v5], [v7, v3], [v1, v4, v6], [v9, v2, v8]]
Graph2: 90 vertices, 188 edges
Greedy chromatic number: 6
\end{lstlisting}
\columnbreak
\includegraphics[width=\columnwidth]{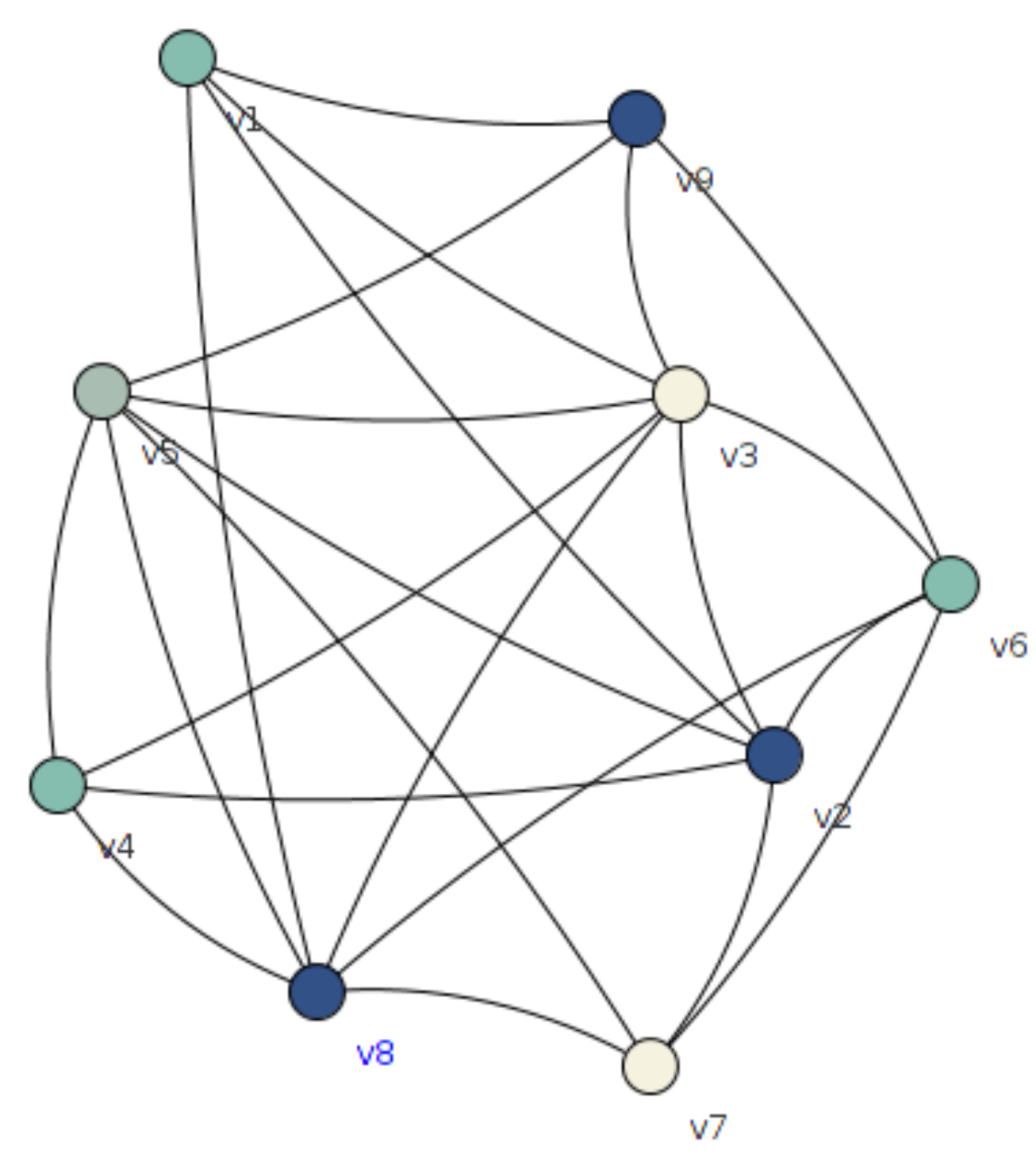}
\includegraphics[width=\columnwidth]{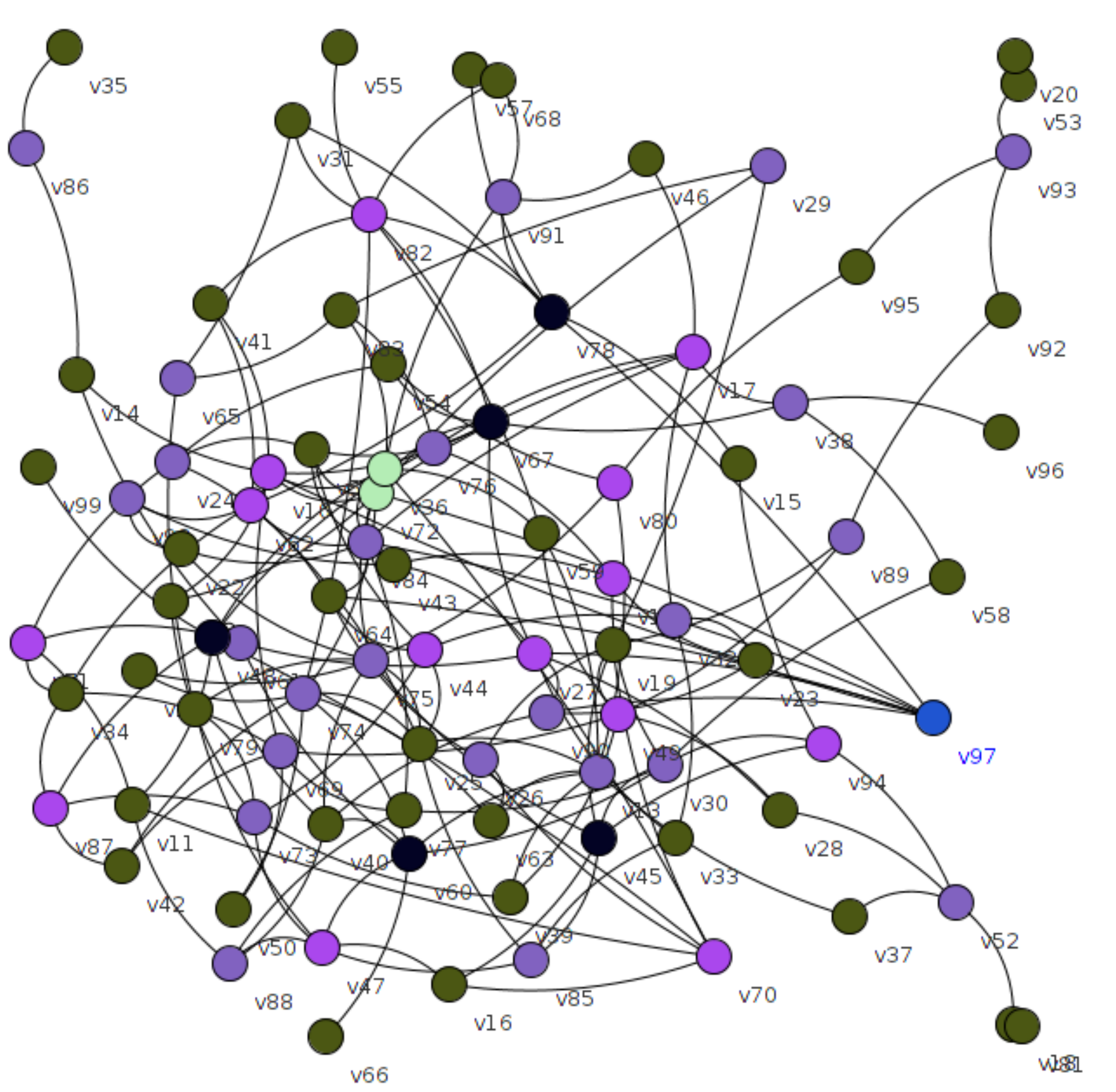}
\end{multicols}

\newpage
\subsubsection{Using MinDFVS algorithms}
\label{tuto:mindfvs}

After generating a graph using one of the factories, this example shows how to compute the minimum directed feedback vertex set.

\begin{multicols}{2}
\begin{lstlisting}[style=java,title=Tutorial: agape.algos.MinDFVS]
/*
 * Copyright University of Orleans - ENSI de Bourges. 
 * Source code under CeCILL license.
 */
package agape.tutorials;

import java.util.ArrayList;
import java.util.Set;

import agape.algos.MinDFVS;
import agape.visu.Visualization;
import edu.uci.ics.jung.graph.DirectedSparseGraph;
import edu.uci.ics.jung.graph.util.Pair;

public class AlgoDFVSTutorial {
    public static void main(String[] args) {

        DirectedGraphFactoryForStringInteger factory = new DirectedGraphFactoryForStringInteger();
        DirectedSparseGraph<String, Integer> g = new DirectedSparseGraph<String, Integer>();
        
        g.addVertex("n1");  g.addVertex("n2");
        g.addVertex("n3");  g.addVertex("n4");
        g.addVertex("n5");  g.addVertex("n6");
        g.addVertex("n7");  g.addVertex("n8");

        g.addEdge(1, new Pair<String>("n1", "n2"));
        g.addEdge(2, new Pair<String>("n1", "n4"));
        g.addEdge(3, new Pair<String>("n2", "n3"));
        g.addEdge(4, new Pair<String>("n3", "n5"));
        g.addEdge(5, new Pair<String>("n5", "n2"));
        g.addEdge(6, new Pair<String>("n5", "n3"));
        g.addEdge(7, new Pair<String>("n4", "n6"));
        g.addEdge(8, new Pair<String>("n4", "n7"));
        g.addEdge(9, new Pair<String>("n8", "n1"));
        g.addEdge(10, new Pair<String>("n7", "n8"));
        g.addEdge(11, new Pair<String>("n3", "n1"));
        
        System.out.println(g);

        MinDFVS<String, Integer> dfvs = new MinDFVS<String, Integer>(factory, factory.edgeFactory);
        
        Set<ArrayList<String>> circuits = dfvs.enumAllCircuitsTarjan(g);
        System.out.println("Circuits: " + circuits);

        Set<String> subset = dfvs.maximumDirectedAcyclicSubset(g);
        System.out.println("Max directed acyclic subset: " + subset);
        
        Set<String> subset3 = dfvs.greedyMinFVS(g);
        System.out.println("Greedy directed acyclic subset:" + subset3);
                
        Set<String> subsetComplementary = dfvs.minimumDirectedAcyclicSubset(g);
        System.out.println("Min directed acyclic subset: " + subsetComplementary);
        
        Visualization.showGraph(g, subsetComplementary);
    }
}
\end{lstlisting}
\columnbreak
\begin{lstlisting}[style=java,title=Tutorial: console output]
Vertices:n1,n5,n4,n3,n2,n8,n7,n6
Edges:1[n1,n2] 2[n1,n4] 3[n2,n3] 4[n3,n5] 5[n5,n2] 6[n5,n3] 7[n4,n6] 8[n4,n7] 9[n8,n1] 10[n7,n8] 11[n3,n1] 
Circuits: [[n5, n3], [n5, n2, n3], [n1, n4, n7, n8], [n1, n2, n3]]
Max directed acyclic subset: [n1, n5, n4, n2, n8, n6]
Greedy directed acyclic subset:[n1, n5]
Min directed acyclic subset: [n3, n7]
\end{lstlisting}

\begin{center}
\includegraphics[width=\columnwidth]{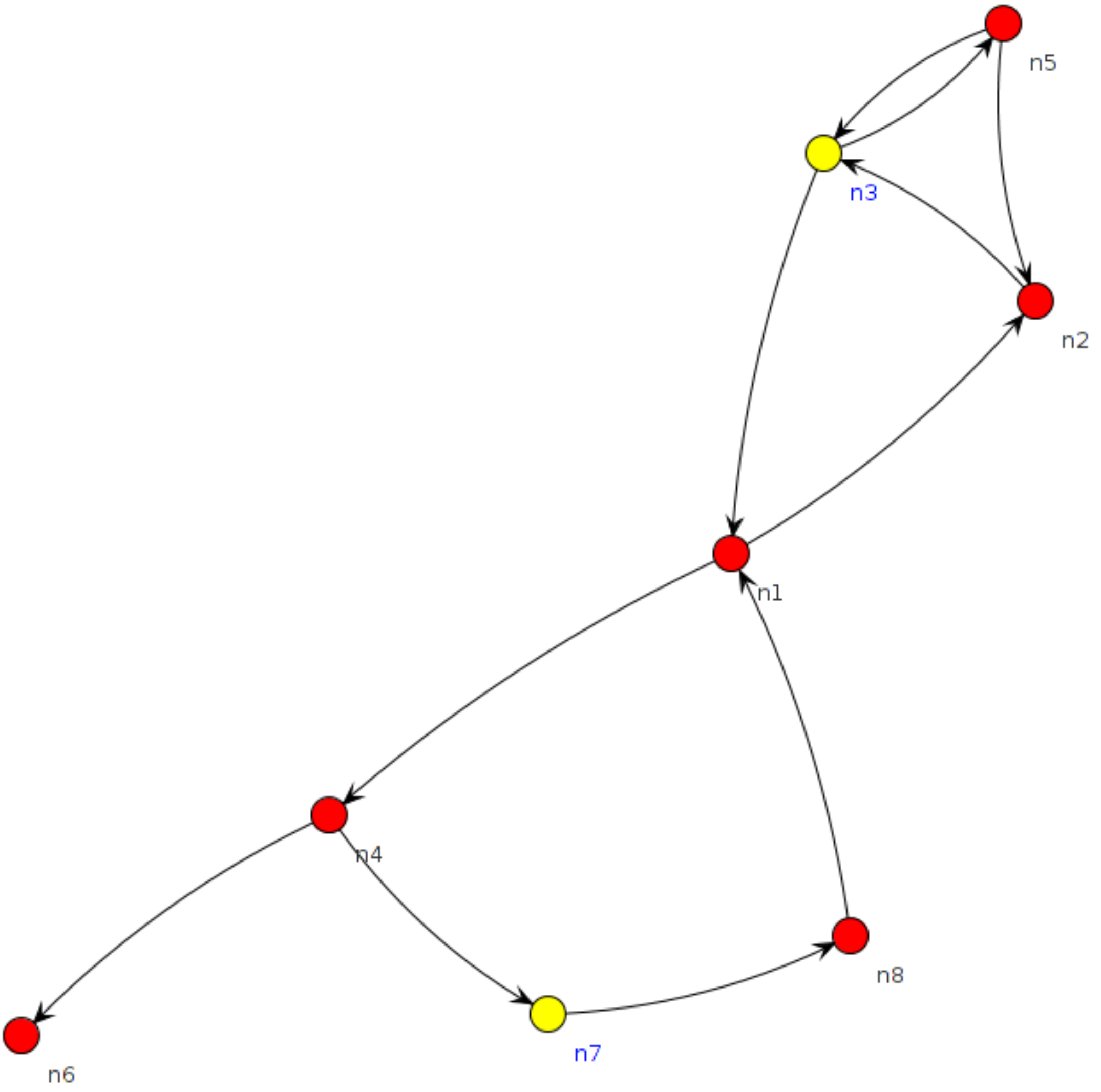}
\end{center}
\end{multicols}

\newpage
\subsubsection{Using MIS algorithms}
\label{tuto:mis}

This example shows to how compute the minimum independent set.

\begin{multicols}{2}
\begin{lstlisting}[style=java,title=Tutorial: agape.algos.MIS]
/*
 * Copyright University of Orleans - ENSI de Bourges. 
 * Source code under CeCILL license.
 */
package agape.tutorials;

import java.util.Set;

import agape.algos.MIS;
import agape.visu.Visualization;
import edu.uci.ics.jung.graph.DirectedSparseGraph;
import edu.uci.ics.jung.graph.util.Pair;

public class AlgoMISTutorial {
    public static void main(String[] args) {

        DirectedGraphFactoryForStringInteger factory = new DirectedGraphFactoryForStringInteger();

        DirectedSparseGraph<String, Integer> g = new DirectedSparseGraph<String, Integer>();
        g.addVertex("n1");  g.addVertex("n2");
        g.addVertex("n3");  g.addVertex("n4");
        g.addVertex("n5");

        g.addEdge(1, new Pair<String>("n1", "n2"));
        g.addEdge(2, new Pair<String>("n1", "n4"));
        g.addEdge(3, new Pair<String>("n2", "n3"));
        g.addEdge(4, new Pair<String>("n3", "n5"));
        g.addEdge(5, new Pair<String>("n5", "n2"));
        g.addEdge(6, new Pair<String>("n5", "n3"));

        System.out.println(g);

        Set<String> set = null;
        MIS<String, Integer> mis = new MIS<String, Integer>(factory, factory.vertexFactory, factory.edgeFactory); 
        
        set = mis.maximalIndependentSetGreedy(g);
        System.out.println("Greedy result: " + set);

        set = mis.maximuRmIndependentSetFominGrandoniKratsch(g);
        System.out.println("FominGrandoniKratsch: " + set);

        set = mis.maximumIndependentSetMaximumDegree(g);
        System.out.println("Branching on Maximum Degree: " + set);

        set = mis.maximumIndependentSetMoonMoser(g);
        System.out.println("Moon Moser result: " + set);

        set = mis.maximumIndependentSetMoonMoserNonRecursive(g);
        System.out.println("Moon Moser (non recursive) result: " + set);
        
        Visualization.showGraph(g, set);
    }
}
\end{lstlisting}
\columnbreak
\begin{lstlisting}[style=java,title=Tutorial: console output]
Vertices:n1,n5,n4,n3,n2
Edges:1[n1,n2] 2[n1,n4] 3[n2,n3] 4[n3,n5] 5[n5,n2] 6[n5,n3] 
Greedy result: [n4, n2]
FominGrandoniKratsch: [n1, n5]
Branching on Maximum Degree: [n1, n5]
Moon Moser result: [n1, n5]
Moon Moser (non recursive) result: [n4, n2]
\end{lstlisting}
\begin{center}
\includegraphics[width=\columnwidth]{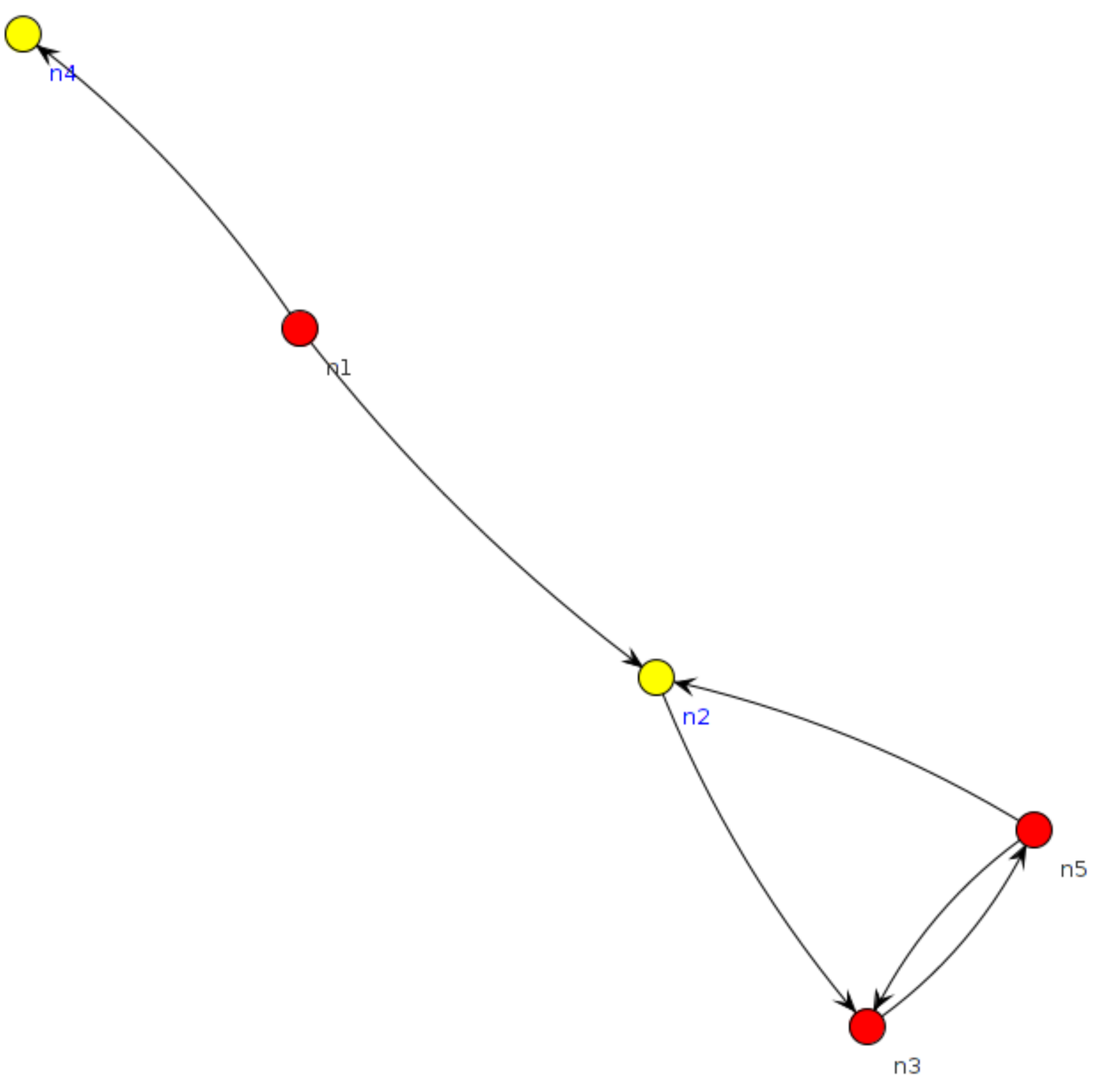}
\end{center}
\end{multicols}

\newpage
\subsubsection{Using MVC algorithms}
\label{tuto:mvc}

This example shows computations of the minimum vertex cover for undirected graphs.

\begin{multicols}{2}
\begin{lstlisting}[style=java,title=Tutorial: agape.algos.MVC]
/*
 * Copyright University of Orleans - ENSI de Bourges. 
 * Source code under CeCILL license.
 */
package agape.tutorials;

import agape.algos.MVC;
import agape.visu.Visualization;
import edu.uci.ics.jung.graph.SparseGraph;
import edu.uci.ics.jung.graph.util.Pair;

public class AlgoMVCTutorial {
    public static void main(String[] args) {

        System.out.println("----------------");
        System.out.println("UNDIRECTED GRAPH");
        System.out.println("----------------");

        SparseGraph<String, Integer> gu = new SparseGraph<String, Integer>();
        UndirectedGraphFactoryForStringInteger undfactory = new UndirectedGraphFactoryForStringInteger();

        gu.addVertex("n1");  gu.addVertex("n2");
        gu.addVertex("n3");  gu.addVertex("n4");
        gu.addVertex("n5");  gu.addVertex("n6");
        gu.addVertex("n7");  gu.addVertex("n8");

        gu.addEdge(1, new Pair<String>("n1", "n2"));
        gu.addEdge(2, new Pair<String>("n1", "n4"));
        gu.addEdge(3, new Pair<String>("n2", "n3"));
        gu.addEdge(5, new Pair<String>("n5", "n2"));
        gu.addEdge(7, new Pair<String>("n1", "n6"));
        gu.addEdge(9, new Pair<String>("n1", "n7"));
        gu.addEdge(11, new Pair<String>("n2", "n8"));

        System.out.println(gu);

        MVC<String, Integer> mvc = new MVC<String, Integer>(undfactory);

        mvc.twoApproximationCover(gu);
        System.out.println("result 2-approx: " + mvc.getVertexCoverSolution());

        mvc.greedyCoverMaxDegree(gu);
        System.out.println("result greedy: " + mvc.getVertexCoverSolution());

        mvc.kVertexCoverBruteForce(gu, 2);
        System.out.println("result BruteForce k=2: " + mvc.getVertexCoverSolution());

        mvc.kVertexCoverBruteForce(gu, 3);
        System.out.println("result BruteForce k=3: " + mvc.getVertexCoverSolution());

        mvc.kVertexCoverBruteForce(gu, 4);
        System.out.println("result BruteForce k=4: " + mvc.getVertexCoverSolution());

        mvc.kVertexCoverDegreeBranchingStrategy(gu, 2);
        System.out.println("result DegreeBranching k=2: " + mvc.getVertexCoverSolution());

        mvc.kVertexCoverBussGoldsmith(gu, 2);
        System.out.println("result Buss k=2: " + mvc.getVertexCoverSolution());

        mvc.kVertexCoverNiedermeier(gu, 2);
        System.out.println("result Niedermeier k=2: " + mvc.getVertexCoverSolution());

        Visualization.showGraph(gu, mvc.getVertexCoverSolution());
    }
}
\end{lstlisting}
\columnbreak
\begin{lstlisting}[style=java,title=Tutorial: console output]
----------------
UNDIRECTED GRAPH
----------------
Vertices:n1,n5,n4,n3,n2,n8,n7,n6
Edges:1[n1,n2] 2[n1,n4] 3[n2,n3] 5[n5,n2] 7[n1,n6] 9[n1,n7] 11[n2,n8] 
result 2-approx: [n1, n2]
result greedy: [n1, n2]
result BruteForce k=2: [n1, n2]
result BruteForce k=3: [n1, n5, n2]
result BruteForce k=4: [n1, n5, n3, n8]
result DegreeBranching k=2: [n1, n2]
result Buss k=2: [n1, n2]
result Niedermeier k=2: [n1, n2]
\end{lstlisting}

\begin{center}
\includegraphics[width=\columnwidth]{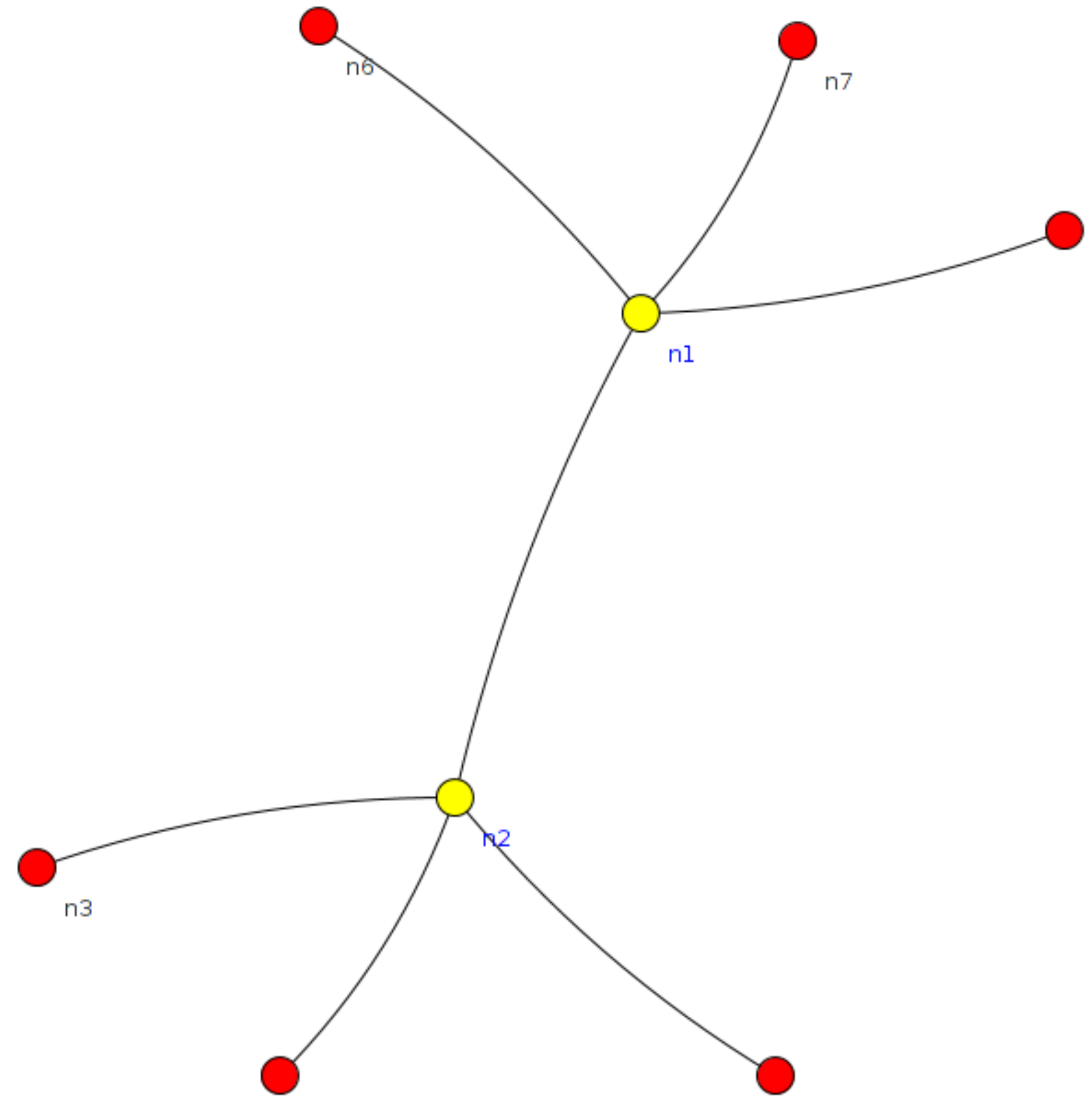}
\end{center}
\end{multicols}

\newpage
\subsubsection{Using Separators algorithms}
\label{tuto:sep}

This example shows how to compute the minimum separators of two vertices.

\begin{multicols}{2}
\begin{lstlisting}[style=java,title=Tutorial: agape.algos.Separators]
/*
 * Copyright University of Orleans - ENSI de Bourges. 
 * Source code under CeCILL license.
 */
package agape.tutorials;

import java.util.HashSet;
import java.util.Set;

import agape.algos.Separators;
import agape.visu.Visualization;
import edu.uci.ics.jung.graph.SparseGraph;
import edu.uci.ics.jung.graph.util.Pair;

public class AlgoSeparatorsTutorial {
    public static void main(String[] args) {

        Separators<String, Integer> sep = new Separators<String, Integer>();
        
        System.out.println("----------------");
        System.out.println("UNDIRECTED GRAPH");
        System.out.println("----------------");
        
        SparseGraph<String, Integer> gu = new SparseGraph<String, Integer>();
        
        gu.addVertex("n1");  gu.addVertex("n2");
        gu.addVertex("n3");  gu.addVertex("n4");
        gu.addVertex("n5");  gu.addVertex("n6");
        
        gu.addEdge(1, new Pair<String>("n1", "n2"));
        gu.addEdge(2, new Pair<String>("n1", "n4"));
        gu.addEdge(3, new Pair<String>("n2", "n3"));
        gu.addEdge(4, new Pair<String>("n3", "n5"));
        gu.addEdge(5, new Pair<String>("n5", "n2"));
        gu.addEdge(6, new Pair<String>("n1", "n6"));
        gu.addEdge(7, new Pair<String>("n5", "n6"));
        
        System.out.println(gu);
                
        Set<Set<String>> set = sep.getABSeparators(gu, "n4", "n3");
        System.out.println("n4/n5: " + set);
        
        set = sep.getAllMinimalSeparators(gu);
        System.out.println("all minimum separators: " + set);
        
        set = sep.getABSeparators(gu, "n1", "n5");
        System.out.println("n1/n5: " + set);
        
        HashSet<String> toSeparate = new HashSet<String>();
        toSeparate.add("n1");
        toSeparate.add("n5");
        Visualization.showGraph(gu, set.iterator().next(), toSeparate);       
    }
}
\end{lstlisting}
\columnbreak
\begin{lstlisting}[style=java,title=Tutorial: console output]
----------------
UNDIRECTED GRAPH
----------------
Vertices:n1,n5,n4,n3,n2,n6
Edges:1[n1,n2] 2[n1,n4] 3[n2,n3] 4[n3,n5] 5[n5,n2] 6[n1,n6] 7[n5,n6] 
n4/n5: [[n1], [n2, n6], [n5, n2]]
all minimum separators: [[n1], [n2, n6], [n5, n2], [n1, n5]]
n1/n5: [[n2, n6]]
\end{lstlisting}
\begin{center}
\includegraphics[width=\columnwidth]{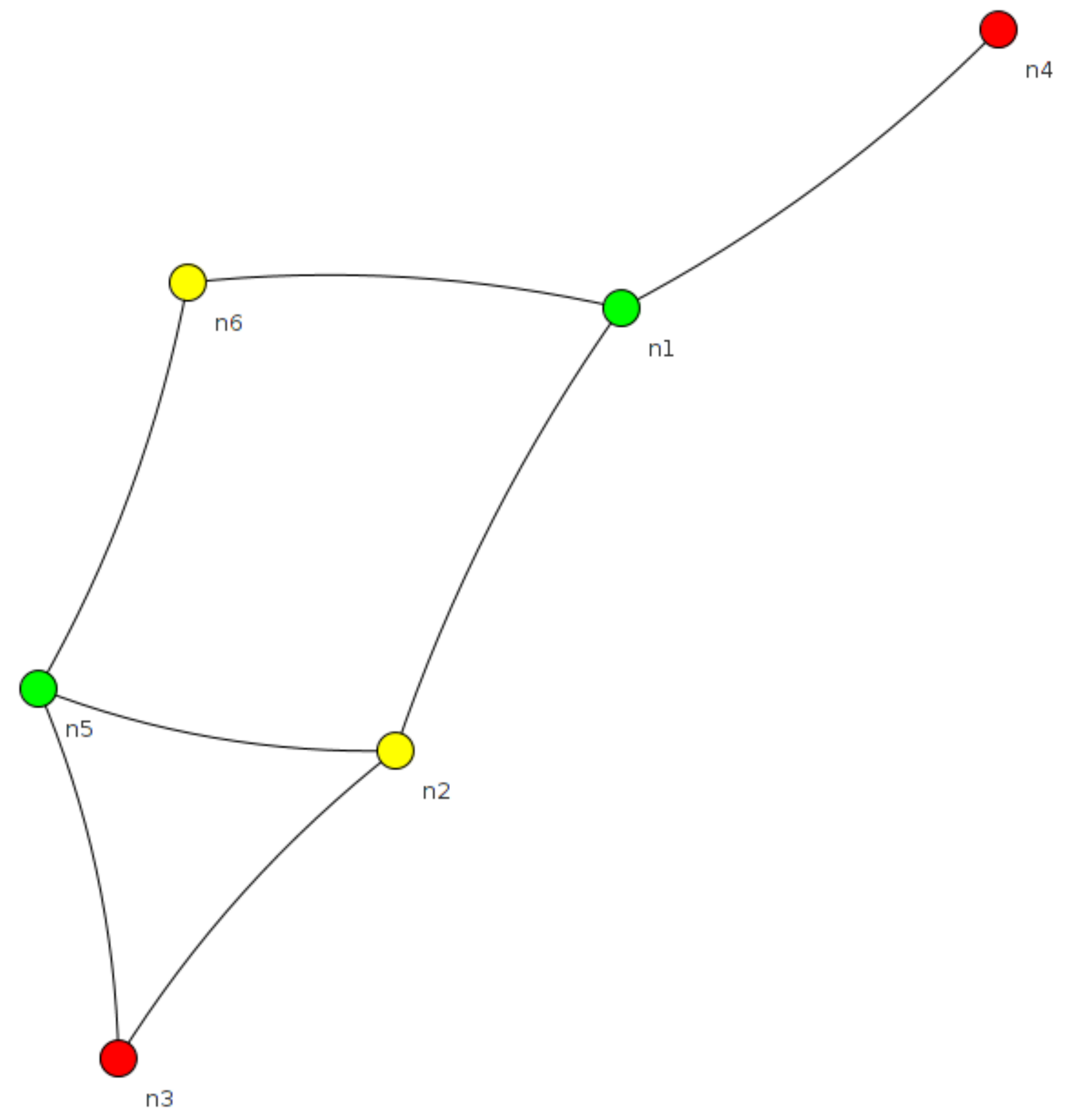}
\end{center}
\end{multicols}

\newpage

\addcontentsline{toc}{section}{References}  
\bibliographystyle{plain} 
\bibliography{biblio}

\end{document}